\documentclass[preprint,12pt]{elsarticle}

\usepackage[margin=1in]{geometry}

\usepackage{graphicx}

\usepackage{amssymb}

\usepackage{url}

\usepackage{amsmath}
\usepackage{upgreek}

\usepackage{amsthm}

\def\doubleunderline#1{\underline{\underline{#1}}}
\usepackage{tensor}
\usepackage{dsfont}

\usepackage{siunitx}

\usepackage{amssymb}

\usepackage{bbm}

\usepackage{soul}

\usepackage{chemformula}

\usepackage{algorithmicx}
\usepackage{algorithm}
\usepackage[noend]{algpseudocode}
\makeatletter
\renewcommand{\ALG@beginalgorithmic}{\footnotesize}
\makeatother

\usepackage{booktabs}

\usepackage{amsmath}

\makeatletter
\newcommand{\vast}{\bBigg@{4}}
\newcommand{\Vast}{\bBigg@{7}}
\makeatother

\makeatletter
\def\ps@pprintTitle{%
  \let\@oddhead\@empty
  \let\@evenhead\@empty
  \let\@oddfoot\@empty
  \let\@evenfoot\@oddfoot
}
\makeatother

\usepackage{todonotes}

\newcommand*\xbar[1]{%
   \hbox{%
     \vbox{%
       \hrule height 0.5pt 
       \kern0.5ex
       \hbox{%
         \kern-0.1em
         \ensuremath{#1}%
         \kern-0.1em
       }%
     }%
   }%
}

\usepackage{caption}

\usepackage{subcaption}

\setcitestyle{authoryear}
 \biboptions{comma,round}

\date{}

\journal{Elsevier}

\begin{document}

\begin{frontmatter}

\title{Accurate conservative phase-field method for simulation of two-phase flows}


\author{Suhas S. Jain\corref{cor1}}
\ead{sjsuresh@stanford.edu}

\address{Center for Turbulence Research, Stanford University, California, USA 94305}

\begin{abstract}

In this work, we propose a novel phase-field model for the simulation of two-phase flows that is accurate, conservative, bounded, and robust. The proposed model conserves the mass of each of the phases, and results in bounded transport of the volume fraction. We present results from the canonical test cases of a drop advection and a drop in a shear flow, showing significant improvement in the accuracy over the commonly used conservative phase-field method.

Moreover, the proposed model imposes a lesser restrictive Courant-Friedrichs-Lewy condition, and hence, is less expensive compared to other conservative phase-field models. We also propose improvements on computation of surface tension forces and show that the proposed improvement significantly reduces the magnitude of spurious velocities at the interface. 





We also derive a consistent and conservative momentum transport equation for the proposed phase-field model and show that the proposed model when coupled with the consistent momentum transport equation results in discrete conservation of kinetic energy, which is a sufficient condition for the numerical stability of incompressible flows, in the absence of dissipative mechanisms. 
To illustrate the robustness of the method in simulating high-density ratio turbulent two-phase flows, we present the numerical simulations of high-density ratio droplet-laden isotropic turbulence at finite and infinite Reynolds numbers.

\end{abstract}

\begin{keyword}
phase-field method \sep two-phase flows \sep turbulent flows \sep ACDI/ACPF


\end{keyword}

\end{frontmatter}



\section{Introduction\label{sec:intro}}

Two-phase flows are ubiquitous in nature and have applications in engineering and natural processes such as atomization of jets and sprays \citep{lin1998drop}, breaking waves \citep{lamarre1991air}, emulsions \citep{bibette1999emulsions}, boiling phenomena \citep{dhir1998boiling}, carbon sequestration \citep{lal2008carbon}, and bubbly flows in cooling towers of nuclear power plants \citep{ha2011development}. 
Numerous challenges are associated with the numerical modeling of two-phase flows, primarily due to the presence of discontinuities at the interface and the inherent multiscale nature of the problem. 
Over the years, many interface-capturing and interface-tracking methods have been proposed. The interface-capturing methods are preferred over interface-tracking methods due to the capability of the interface-capturing methods to handle dynamic topological changes implicitly. These methods can be broadly classified into a volume-of-fluid (VOF) method, a level-set (LS) method, a phase-field (PF) method, and hybrid methods such as a CLSVOF (coupled level-set and volume-of-fluid) method, and a VOSET (coupled volume-of-fluid and level-set) method, where typically two methods are combined to have the advantages of both methods \citep{Mirjalili2017}. 
The VOF and LS methods are sharp-interface methods and are generally known to be accurate in representing the material interface, which is physically sharp in the continuum limit. 
The main disadvantages of these methods are that a VOF method involves an interface-reconstruction step that is complex and expensive, and an LS method is not conservative. The hybrid methods were proposed to overcome some of these challenges and combine the advantages of two methods, but they are known to be more complicated than the original methods.

A PF method, in contrast, is a diffuse-interface approach that is relatively less expensive and simple to implement. Some PF models are also inherently conservative. Hence, we adopt a diffuse-interface approach in this work, and propose a novel PF method that is conservative and more accurate than some existing PF methods in the literature, and we refer to this new model as an accurate conservative diffuse-interface/phase-field (ACDI/ACPF) model.


\section{Phase-field method}

\subsection{Cahn-Hilliard and Allen-Cahn models}

The classical PF methods are based on Cahn-Hilliard and Allen-Cahn equations \citep{cahn1958free,allen1979microscopic} that were originally developed to model the phase separation and coarsening phenomena in solids and the motion of antiphase boundaries in crystalline solids, respectively. More recently, these methods have been adopted for modeling the interface between two fluids \citep{anderson1998diffuse, lowengrub1998quasi, chen1998applications, jacqmin1999calculation,liu2003phase,biben2003tumbling,badalassi2003computation,yue2004diffuse,yang2006numerical,kim2012phase,lamorgese2017modeling,soligo2019mass}. A Cahn-Hilliard PF model is conservative but involves a fourth-order spatial derivative in the equation, which requires careful construction of the numerical methods. In contrast, an Allen-Cahn PF model does not involve fourth-order derivatives in the equation, but is not conservative. 

\subsection{Conservative phase-field model}

Starting from the Allen-Cahn equation, and subtracting the curvature-driven motion of the interface, \citet{chiu2011conservative} derived a conservative diffuse-interface/phase-field (CDI/CPF) model for incompressible flows, which can be written as
\begin{equation}
\frac{\partial \phi}{\partial t} + \vec{\nabla}\cdot(\vec{u}\phi) = \vec{\nabla}\cdot\left\{\Gamma\left[\epsilon\vec{\nabla}\phi - \phi(1 - \phi)\frac{\vec{\nabla} \phi}{|\vec{\nabla} \phi|}\right]\right\},
\label{eq:CDI}
\end{equation}
where $\phi$ is the PF variable which represents the volume fraction, $\vec{u}$ is the velocity, $\Gamma$ represents the velocity-scale parameter, and $\epsilon$ is the interface thickness scale parameter. 
This model can be thought of as the one-step conservative LS method introduced by \citet{olsson2005conservative}, and is also sometimes referred to as a conservative/modified Allen-Cahn equation because it can be derived analytically starting from the Allen-Cahn equation. This model has since been widely used not only in the finite-volume/finite-difference setting but also in lattice-Boltzmann methods \citep{geier2015conservative,wang2016comparative,ren2016improved,liang2018phase,fakhari2018phase,abadi2018conservative,aihara2019multi}. Using central difference schemes, \citet{Mirjalili2020} showed that the PF variable, $\phi$, remains bounded between $0$ and $1$ for incompressible flows, if the values for $\Gamma$ and $\epsilon$ are appropriately chosen. More recently, \citet{jain2020conservative} proposed a CDI method for compressible flows and showed that $\phi$ remains bounded between $0$ and $1$ for compressible flows, with the use of central difference schemes, provided that an additional constraint on time-step size is satisfied. Additionally, \citet{jain2020conservative} showed that the transport of $\phi$ with this PF method also satisfies the total-variation diminishing (TVD) property. The compressible CDI model has been further coupled with a shock-capturing method for the simulation of high-Mach-number two-phase flows, extended to multiphase modeling of deforming solids, and also extended for higher-order explicit and compact central schemes by \citet{jain2021assessment}.

\begin{figure}
    \centering
    \includegraphics[width=0.4\textwidth]{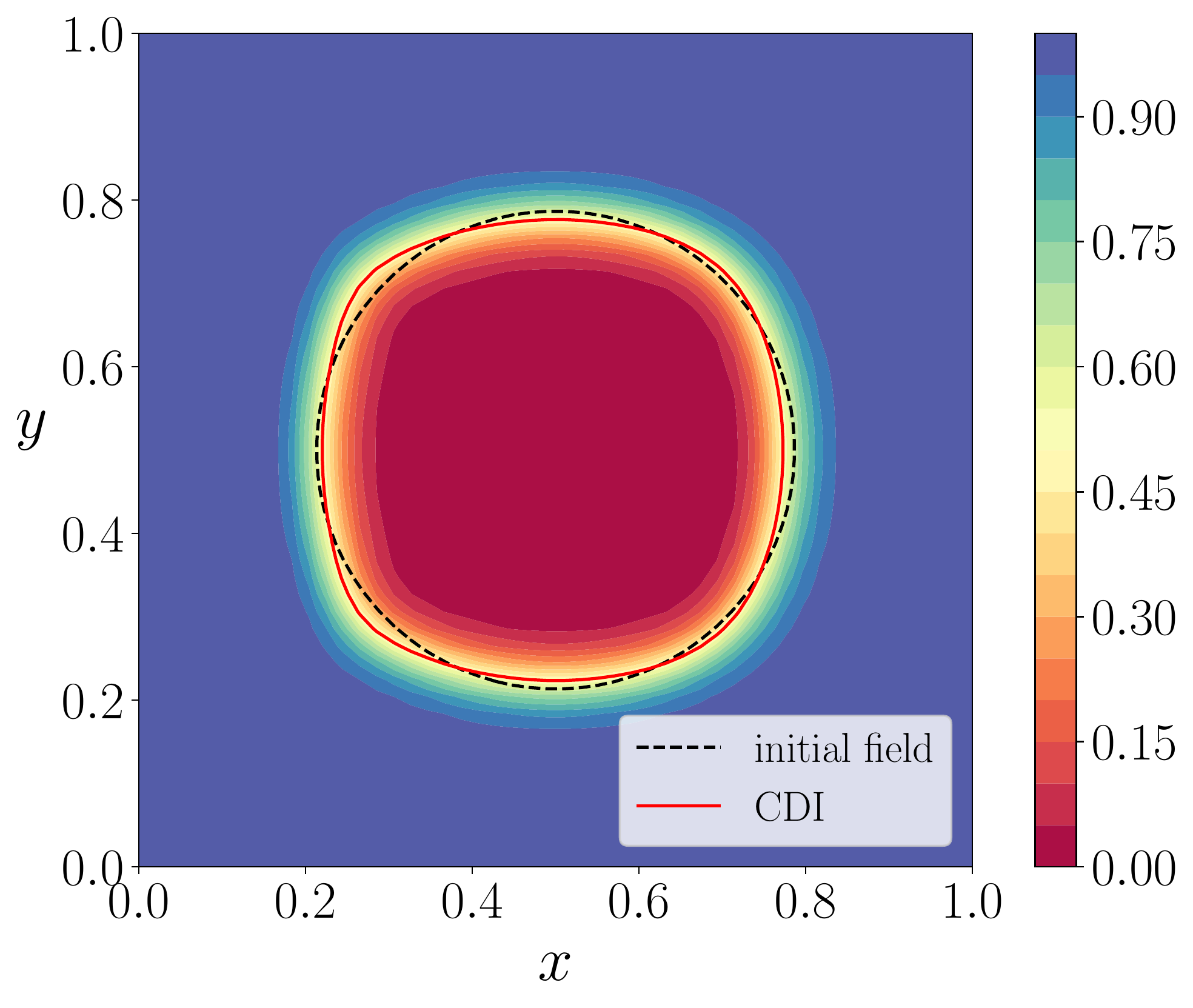}
    \caption{Final shape of the translating bubble/drop after $5$ flow through times, computed using Eq. \eqref{eq:CDI}, illustrating the shape distortion of the interface.}
    \label{fig:distort}
\end{figure}

This CDI/CPF method is easy to implement and cost-effective, but the two main issues with this model are (a) artificial distortion of the interface (alignment with the grid) and (b) limited accuracy. 
\citet{jain2021assessment} compared the compressible CDI method of \citet{jain2020conservative} with the quasi-conservative diffuse-interface method by \citet{shukla2010interface} and \citet{tiwari2013diffuse}, and with the localized-artificial diffusivity based diffuse-interface method by \cite{cook2007artificial}, \citet{subramaniam2018high}, and \citet{adler2019strain} in the context of compressible flows and made the following conclusions. The CDI method is superior in maintaining the conservation property and in maintaining a constant interface thickness throughout the simulation, but it results in artificial distortion of the interface for long-time integrations and in the absence of surface tension effects. This behavior is, however, not limited to compressible flows and is valid for both incompressible and compressible flows. 
To reproduce this interface distortion behavior seen before with the CDI method, we here present the simulation results of advection of a two-dimensional circular drop/bubble in a periodic domain by solving only the CDI phase-field equation in Eq. \eqref{eq:CDI}  (decoupled from hydrodynamics) with a prescribed velocity field of $\vec{u}=5i$, {where $i$ is the unit vector along the x direction}. The domain has the size of $1\times 1$ and is discretized into $50\times50$ grid points. The final and initial shapes of the drop, after $5$ flow-through times at time $t=1$, are shown in Figure \ref{fig:distort}. The shape of the drop can be seen to be significantly distorted from the original circular shape at time $t=0$. 
This alignment of the interface with the grid was also observed by \citet{tiwari2013diffuse} with the quasi-conservative diffuse-interface method and by \citet{Waclawczyk2015} and \citet{chiodi2017reformulation} in the context of a conservative level-set method. Note that, the exact amount of interface distortion is probably dependent on the choice of the numerical discretization. 
Some fixes proposed for this issue are to use a higher-order scheme \citep{tiwari2013diffuse,jain2021assessment} and the reformulation of the reinitialization procedure in the conservative LS method \citep{chiodi2017reformulation}. 

To maintain the boundedness of $\phi$ with the CDI method, the criterion that needs to be satisfied \citep{Mirjalili2020,jain2020conservative} is
\begin{equation}
    \Gamma^* \ge \frac{1}{(2\epsilon^* - 1)},
    \label{eq:CDI_bound}
\end{equation}
where $\Gamma^*=\Gamma/|\vec{u}|_{max}$ and $\epsilon^*=\epsilon/(\Delta x)$ are the non-dimensionalized parameters.
One way to improve the accuracy of a diffuse-interface method without increasing the grid resolution is by reducing $\epsilon$, which improves the accuracy by making the interface sharper. The discussions on the sharp-interface limit of Allen-Cahn based phase-field models can be found in \citet{zhang2009numerical} and \citet{abels2022sharp}. But with the CDI method, as $\epsilon$ decreases, $\Gamma$ needs to increase to keep $\phi$ bounded. For example, if $\epsilon$ is to be made smaller than one grid size ($\Delta x$), then $\Gamma>|\vec{u}|_{max}$. This increase in the value of $\Gamma$ degrades the accuracy of the model and further increases the cost by imposing a more severe Courant-Friedrich-Lewy (CFL) restriction. Hence, the accuracy of the CDI model degrades as $\epsilon$ becomes small, particularly as $\epsilon$ approaches $0.5\Delta x$, which is not ideal. More recently, \citet{chiu2019coupled} proposed an alternate model with a hope to obtain a model that is more accurate compared to the CDI model. But their proposed model was not conservative, and required solving two transport equations.  

Being aware of all these issues associated with the PF methods in the literature, we therefore seek a method that is
more accurate than the CDI method,
is conservative,
maintains boundedness of the volume fraction,
converges (as $\epsilon$ decreases),
is cost-effective, and is easy to implement and robust.
With this objective, in this work, we propose a novel PF model, which we refer to as an accurate conservative diffuse-interface/phase-field (ACDI/ACPF) method, that has all the above-listed characteristics.
Furthermore, following previous approaches \citep{abels2012thermodynamically,Raessi2012, LeChenadec2013, huang2020allen, huang2020cahn, MIRJALILI2021consistent} 
we derive a consistent and conservative momentum transport equation for the proposed PF model, such that the resulting transport of momentum is consistent with the transport of the two-phase mixture mass and results in discrete kinetic energy conservation in the absence of dissipative mechanisms.

\section{Proposed accurate conservative phase-field model\label{sec:ACDI}}

\begin{figure}
    \centering
    \includegraphics[width=0.5\textwidth]{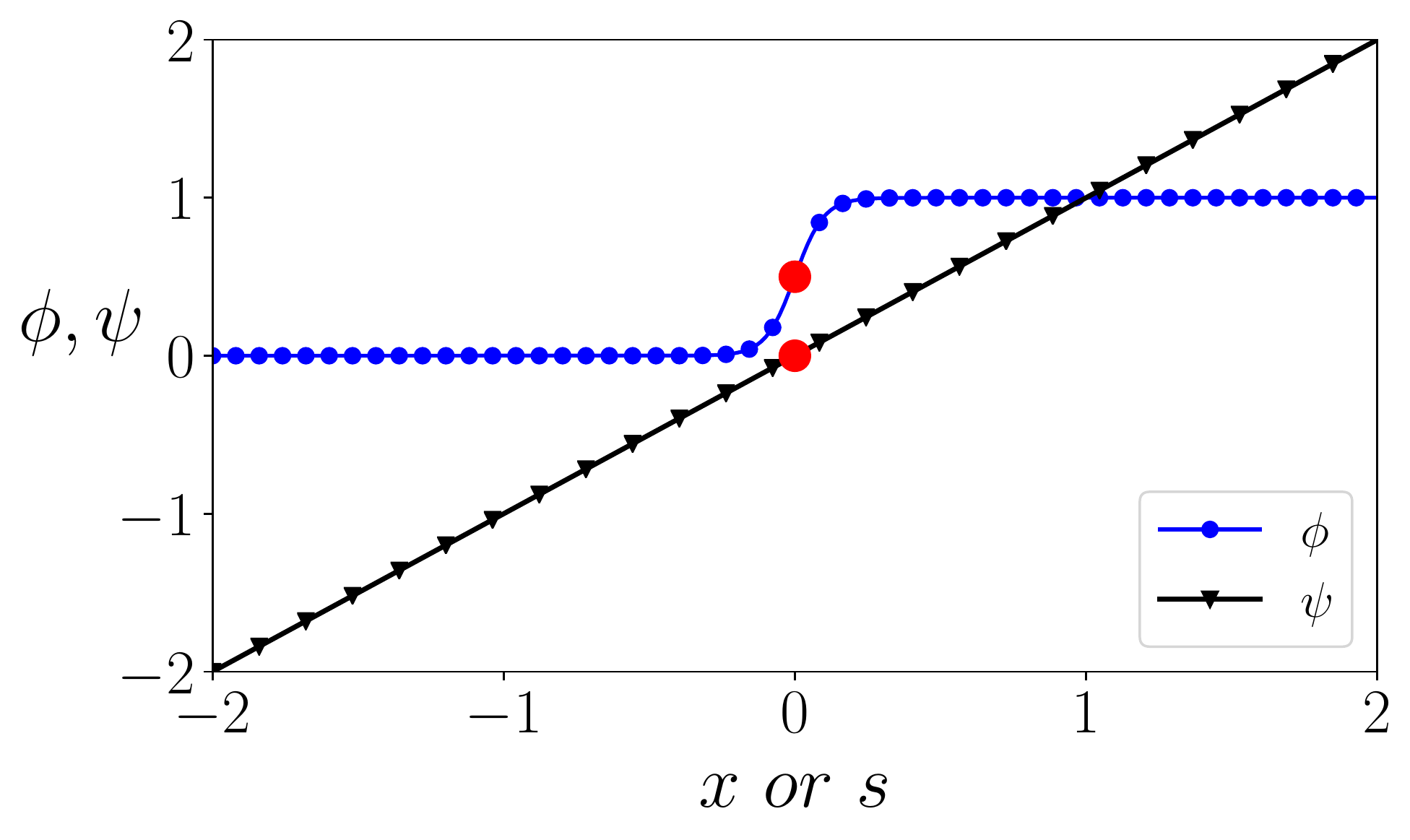}
    \caption{An illustration of the PF variable, $\phi$, and the signed-distance function, $\psi$, for an interface located at $x=0$. The two large dots are marked on the plots at $\phi=0.5$ and $\psi=0$, representing the location of the interface.}
    \label{fig:phi_psi}
\end{figure}

Ideally, we need the interface to be in an equilibrium state. This is achieved when the right-hand-side (RHS) terms in Eq. \eqref{eq:CDI}\textemdash also called interface-regularization terms\textemdash are the stiffest terms. In this limit, 
\begin{equation*}
\vec{\nabla}\cdot\Big[\epsilon\vec{\nabla}\phi - \phi(1 - \phi)\vec{n}\Big] = 0.
\end{equation*}
If $s$ represents a coordinate along the interface normal, then the above equation can now be rewritten along this coordinate, with the normal $\vec{n}=\pm1$, as
\begin{equation*}
\frac{d^2 \phi}{d s^2} \pm \frac{1}{\epsilon} \frac{d}{d s}\left[\phi (1-\phi) \right] = 0.
\end{equation*}
Choosing $\vec{n}=+1$, integrating the above relation, and using the conditions $\phi\rightarrow0$ as $s\rightarrow-\infty$, we arrive at the relation 
\begin{equation*}
    \frac{d \phi}{d s} = \frac{\phi(1 - \phi)}{\epsilon}.
    \label{eq:condition1}
\end{equation*}
Further, let $\psi$ represent the signed-distance function from the interface, such that $\psi$ and $s$ are related as $\psi (\phi)=s(\phi) - s(\phi=0.5)$. With this, the above relation can be re-expressed in terms of $\psi$ as
\begin{equation}
    \frac{d \phi}{d \psi} = \frac{\phi(1 - \phi)}{\epsilon}.
    \label{equ:condition}
\end{equation}
Now, integrating this relation and using the conditions $\phi=0.5$ at $\psi=0$, we arrive at the equilibrium kernel function for $\phi$ in terms of the signed-distance function $\psi$ as
\begin{equation}
\phi = \frac{e^{(\psi/\epsilon)}}{1 + e^{(\psi/\epsilon)}} = \frac{1}{2} \left[1 + \tanh{\left(\frac{\psi}{2\epsilon}\right)}\right].
\label{equ:phi_sol}
\end{equation}
To illustrate, a plot of $\phi$ and $\psi$, along $s$, is shown in Figure \ref{fig:phi_psi} for an interface located at $x=0$.
Now, using the relations in Eqs. \eqref{equ:condition}-\eqref{equ:phi_sol}, we can arrive at the relation
\begin{equation}
    \phi(1 - \phi) = \frac{1}{4} \left[1 - \tanh^2{\left(\frac{\psi}{2\epsilon}\right)}\right].
    \label{equ:identity}
\end{equation}
Further, we can also show that the normal can be reexpressed in terms of $\psi$ as
\begin{equation}
    \vec{n}=\frac{\vec{\nabla} \phi}{|\vec{\nabla} \phi|} = \frac{\vec{\nabla} \psi}{|\vec{\nabla} \psi|}.
    \label{eq:normal}
\end{equation}
Similar relations were used by \citet{Waclawczyk2015}, \citet{chiodi2017reformulation}, and \citet{chiu2019coupled} to reformulate interface-capturing methods.
Using the relation in Eq. \eqref{equ:identity}, we replace the non-linear sharpening flux term $\phi(1-\phi)$ in Eq. \eqref{eq:CDI} with the expression in terms of $\psi$. As shown in Figure \ref{fig:phi_psi}, $\psi$ is a better-behaved function compared with $\phi$ that does not contain any jumps/discontinuities. Hence, using $\psi$ to compute the non-linear sharpening flux will result in more accurate discrete representation of the flux. Furthermore, with the same reasoning, we also replace the computation of the normal term $\vec{\nabla} \phi/|\vec{\nabla} \phi|$ in Eq. \eqref{eq:CDI} with the expression in terms of $\psi$ in Eq. \eqref{eq:normal}. With these modifications, the proposed new PF model, an ACDI/ACPF method, can be written as
\begin{equation}
\frac{\partial \phi}{\partial t} + \vec{\nabla}\cdot(\vec{u}\phi) = \vec{\nabla}\cdot\left\{\Gamma\left\{\epsilon\vec{\nabla}\phi - \frac{1}{4} \left[1 - \tanh^2{\left(\frac{\psi}{2\epsilon}\right)}\right]\frac{\vec{\nabla} \psi}{|\vec{\nabla} \psi|}\right\}\right\}.
\label{eq:ACDI}
\end{equation}

Not only is this form of the PF model derived to obtain a method that is more accurate than the CDI method, but also the terms in the model are carefully constructed such that the model also results in the transport of $\phi$ that is always bounded between $0$ and $1$ (see Section \ref{sec:bound}). For example, the diffusion term in the proposed ACDI model in Eq. \eqref{eq:ACDI} is still written in terms of $\phi$, and we do not replace this with an equivalent expression for $\psi$. This is because, in the limit of $\phi\rightarrow 0\ or\ 1$, the diffusion term that is expressed in terms of $\psi$ could switch signs due to the error in the computation of $\psi$. This will result in failure of the method to maintain a bounded $\phi$ between $0$ and $1$, and can potentially make the method non-robust.

All the terms in the proposed model are in divergence form, and hence, the model is conservative {(conserves the mass of each phase)}. The proposed model is also cost-effective and easy to implement because it requires solving a single scalar partial-differential equation. Additionally, the superior robustness and the convergence  (with $\epsilon$) properties of the model will be illustrated in the following sections. 



%
\subsection{Calculation of $\psi$ \label{sec:psi}}

The proposed PF model requires the computation of $\psi$, and this can be easily calculated using the algebraic relation
\begin{equation}
    \psi = \epsilon \ln\left(\frac{\phi + \varepsilon}{1 - \phi + \varepsilon}\right),
    \label{eq:psi}
\end{equation}
which is obtained by rearranging the relation in Eq. \eqref{equ:phi_sol}. Here, $\varepsilon$ is a small number
\footnote{If one is using an Intel compiler, it might be beneficial to turn off optimization for the floating-point operations using the flag ``-fp-model precise"} 
that is added, to both the numerator and the denominator, to avoid $\psi$ going to $-\infty$ or $\infty$ when $\phi$ goes to $0$ or $1$, respectively (a value of $\varepsilon=10^{-100}$ is used in this work). 
Additionally, if $\phi$ assumes values outside the range $[0,1]$, the computation of $\psi$ using the relation in Eq. \eqref{eq:psi} results in physically unrealizable values. But, because $\phi$ is always guaranteed to be bounded between $0$ and $1$, with the proposed PF model, the direct computation of $\psi$ using Eq. \eqref{eq:psi} will not be an issue. If one is using a different numerical method where $\phi$ is not guaranteed to be bounded between $0$ and $1$, then one could replace $\phi$ with $\tilde{\phi}={\min(\max(\phi,0),1)}$ in Eq. \ref{eq:psi} to make this step even more robust, but this is not a necessity for the choice of numerical method in this work.
Note that, $\psi$ is only an approximation to a signed-distance function and not a perfect signed-distance function because the interface is generally not in equilibrium (a hyperbolic tangent function) due to the velocity gradients in the flow which pushes the interface away from its equilibrium state. Hence, it is not assumed, anywhere in this work, that $\psi$ is a signed-distance function, e.g., for the computation of normal using Eq. \eqref{eq:normal}, $|\vec{\nabla} \psi|$ is not set to $1$.  


\subsection{Choosing the parameters $\Gamma$ and $\epsilon$ \label{sec:bound}}

The values of $\Gamma$ and $\epsilon$ are chosen such that the transport of $\phi$ is always bounded between $0$ and $1$. The criterion that needs to be satisfied is
\begin{equation}
    \Gamma \ge |\vec{u}|_{max}\ \mathrm{and}\ \epsilon > 0.5 \Delta x,
    \label{eq:bound_crit}
\end{equation}
where $|\vec{u}|_{max}$ is the maximum value of the absolute velocity in the domain, and $\Delta x$ is the grid-cell size. 

It is almost always practical and preferable to choose the minimum possible values for $\Gamma$ and $\epsilon$ that satisfy the criterion in Eq. \eqref{eq:bound_crit}.
Choosing the value of $\Gamma$ to be equal to $|\vec{u}|_{max}$ makes the RHS of Eq. \eqref{eq:ACDI} the stiffest term in the equation, which is required for the interface to maintain an equilibrium kernel shape\textemdash a hyperbolic tangent function\textemdash throughout the simulation. A higher value for $\Gamma$, although it forces the interface to assume the equilibrium kernel shape more quickly, makes the RHS term in Eq. \eqref{eq:ACDI} stiffer and imposes an additional CFL restriction.
Choosing the value of $\epsilon$ to be approximately equal to $0.5 \Delta x$ is also preferred, because this keeps the interface, which is resolved on the grid, as sharp as possible. Increasing $\epsilon$ increases the interface thickness and, hence, reduces the accuracy of the method.

\subsection{Cost estimates and accuracy \label{sec:cost}}

The choice of parameters $\Gamma$ and $\epsilon$ in a diffuse-interface method imposes a restriction on the time-step size due to the viscous CFL criterion, which can be expressed as
\begin{equation}
    \Delta t_{max} \sim \frac{\Delta x^2}{\Gamma \epsilon}. 
\end{equation}
From the boundedness criterion, for the CDI method, in Eq. \eqref{eq:CDI_bound}, as $\epsilon^*\rightarrow0.5$, 
the required $\Gamma^*$ goes to $\infty$.
Because both the proposed ACDI method in Eq. \eqref{eq:ACDI} and the CDI method in Eq. \eqref{eq:CDI} maintain boundedness of $\phi$ for the choice of $\Gamma^*=1$ and $\epsilon^*=1$, the cost of both the CDI and the ACDI methods in terms of the required number of time steps is the same. But because the resolved interface becomes sharper as $\epsilon$ reduces, we are interested in estimating the cost of these methods for $\epsilon^*<1$.

Using the boundedness criterion in Eq. \eqref{eq:bound_crit} and expressing $\Delta t_{max}$ in terms of $\epsilon^*$, for the proposed ACDI method, the restriction on the time-step size is
\begin{equation*}
    \Delta t_{max} \sim \frac{\Delta x}{|\vec{u}|_{max}\epsilon^*}. 
\end{equation*}
If $\epsilon^*$ is chosen to be equal to $0.55$, then $\left.\Delta t_{max}\right\vert_{\epsilon^*=0.55}=20/11\left.\Delta t_{max}\right\vert_{\epsilon^*=1}$, and $N_{\epsilon^*=0.55}=0.55N_{\epsilon^*=1}$, where $N$ is the number of time steps. Therefore, the cost is $1.8$ times lower than that of the simulation with $\epsilon^*=1$. Similarly, if $\epsilon^*=0.51$, then the cost is $1.96$ times lower. In contrast, for the CDI method, $\Delta t_{max}$ can be expressed in terms of $\epsilon^*$ as
\begin{equation*}
    \Delta t_{max} \sim \frac{\Delta x}{|\vec{u}|_{max}} \left(2 - \frac{1}{\epsilon^*}\right). 
\end{equation*}
If $\epsilon^*$ is chosen to be equal to $0.55$, then $\left.\Delta t_{max}\right\vert_{\epsilon^*=0.55}=20/11\left.\Delta t_{max}\right\vert_{\epsilon^*=1}$, and $N_{\epsilon^*=0.55}=5.5N_{\epsilon^*=1}$. Therefore, the cost is $5.5$ times higher than that of the simulation with $\epsilon^*=1$. Similarly, if $\epsilon^*=0.51$, then the cost is $25.5$ times higher. Hence, with the proposed ACDI method, reducing $\epsilon$ not only makes the interface more sharp, but also reduces the cost of the simulation. In contrast, with the CDI method, the cost of the simulation increases as $\epsilon$ reduces. As $\epsilon^*\rightarrow 0.5$, the CDI method becomes prohibitively expensive.

Apart from the cost, reducing $\epsilon$ also increases the accuracy of the proposed ACDI method by making the resolved interface sharper. In contrast, for the CDI method, reducing $\epsilon$, particularly for the values of $\epsilon^*<1$, reduces the accuracy of the method, albeit making the interface sharper. This is because a value of $\epsilon^*<1$ requires $\Gamma^*>1$, which makes the RHS term in Eq. \eqref{eq:CDI} stiffer. This increases the error associated with the calculation of the non-linear flux and the normal (as described in Section \ref{sec:ACDI}) and, therefore, exacerbates the issue of artificial interface distortion that was observed by \citet{tiwari2013diffuse} and \citet{jain2021assessment}. A quantitative comparison of the accuracy of the ACDI and CDI methods, with reducing $\epsilon$, is presented in Section \ref{sec:effect_of_eps}.

\subsection{Improvements in computation of surface tension force\label{sec:surface}}

Surface tension force can be modeled as a stress or a body force. These are called integral and volumetric formulations, respectively. It is crucial to have a discrete balance between the surface tension force and the pressure terms to minimize the spurious currents around the interface \citep{francois2006balanced}. 
The continuum-surface force (CSF) formulation \citep{brackbill1992continuum} is used in this work because it is easy to attain this balance with this formulation.

For a diffuse-interface method, the CSF formulation can be written as $\vec{F}=\sigma \kappa \vec{\nabla} \phi$, where $\kappa$ is the curvature, and can be calculated as
\begin{equation}
    \kappa = -\vec{\nabla}\cdot \vec{n} = -\vec{\nabla} \cdot \left( \frac{\vec{\nabla} \phi}{|\vec{\nabla} \phi|}  \right).
    \label{eq:old_kappa}
\end{equation}
It is known that the error associated with the computation of curvature is the primary reason for the formation of spurious currents \citep{popinet2018numerical}. With the proposed ACDI method, we now have access to the well-behaved, signed-distance function $\psi$. Hence, utilizing the relation in Eq. \eqref{eq:normal}, curvature can instead be calculated as
\begin{equation}
    \kappa = -\vec{\nabla}\cdot \vec{n} = -\vec{\nabla} \cdot \left( \frac{\vec{\nabla} \psi}{|\vec{\nabla} \psi|}  \right),
    \label{eq:new_kappa}
\end{equation}
which reduces the truncation errors in computation of $\kappa$ (Appendix A) and, therefore, the amount of spurious currents. 


\section{Coupling with incompressible Navier-Stokes equations \label{sec:incomp}}

When an interface-capturing method is coupled with the Navier-Stokes equations, it is important to solve a form of the momentum equation that transports momentum consistently with the transport of mass. This is required to make the resulting coupled system robust in the limit of high-density ratio and high-$Re$ flows. 

If we represent the diffusion and sharpening fluxes on the RHS of Eq. \eqref{eq:ACDI}, for phase $l$, as
\begin{equation}
    \vec{a}_l = \Gamma\left\{\epsilon\vec{\nabla}\phi_l - \frac{1}{4} \left[1 - \tanh^2{\left(\frac{\psi}{2\epsilon}\right)}\right]\frac{\vec{\nabla} \psi}{|\vec{\nabla} \psi|}\right\},
    \label{eq:regularization}
\end{equation}
where $\phi_l$ is the volume fraction of phase $l$, then this relation satisfies the condition
\begin{equation}
    \sum_{l=1}^2 \vec{a}_l = 0. 
\end{equation}
The implied mass transport equation for the proposed ACDI method can then be written as
\begin{equation}
\frac{\partial \rho}{\partial t} + \vec{\nabla}\cdot(\vec{u}\rho) = \vec{\nabla}\cdot\vec{f},
\label{eq:mass_incomp}    
\end{equation}
where 
\begin{equation*}
\vec{f}=\sum_{l=1}^2 \rho_l \vec{a}_l = \left\{(\rho_1 - \rho_2) \Gamma\left\{\epsilon\vec{\nabla}\phi_1 - \frac{1}{4} \left[1 - \tanh^2{\left(\frac{\psi}{2\epsilon}\right)}\right]\frac{\vec{\nabla} \psi}{|\vec{\nabla} \psi|}\right\}\right\}.
\end{equation*}

\subsection{Consistent momentum transport equation}

The consistency corrections for the momentum equation for the proposed ACDI method can be derived with the premise that a consistent momentum transport equation does not spuriously contribute to the kinetic energy of the system. 
The full coupled system of equations for the simulation of incompressible two-phase flows can then be written as
\begin{equation}
\frac{\partial \phi_1}{\partial t} + \vec{\nabla}\cdot(\vec{u}\phi_1) = \vec{\nabla}\cdot\vec{a}_1,
\label{eq:volumef}    
\end{equation}
\begin{equation}
\frac{\partial \rho\vec{u}}{\partial t} + \vec{\nabla}\cdot(\rho \vec{u} \otimes \vec{u} + p \mathds{1}) = \vec{\nabla}\cdot\doubleunderline\tau + \vec{\nabla}\cdot(\vec{f}\otimes\vec{u}) + \sigma \kappa \vec{\nabla} \phi_1 + \rho \vec{g},
\label{eq:momf}
\end{equation}
\begin{equation}
\vec{\nabla}\cdot \vec{u} = 0,
\end{equation}
where $\doubleunderline\tau$ is the stress tensor, and $\vec{g}$ represents a generic body force. A fractional-step method \citep{kim1985application} is used to compute the pressure, which closes the system of equations. The resulting kinetic energy transport equation can be written as
\begin{equation}
    \frac{\partial \rho k}{\partial t} + \vec{\nabla}\cdot(\rho \vec{u}k) + \vec{\nabla}\cdot(\vec{u}p) = \vec{\nabla}\cdot(\vec{f}k),
\end{equation}
where all the terms are in conservative form. Thus, the proposed consistent momentum equation [Eq. \eqref{eq:momf}] results in conservation of total kinetic energy. {A discussion on the coupling of the proposed ACDI method with compressible Navier-Stokes equations is presented in Appendix B.}

\section{Numerical method: skew-symmetric-like splitting}

In this work, the equations are discretized on an Eulerian Cartesian grid. A fourth-order Runge-Kutta method is used for time discretization, and a second-order flux-split conservative finite-difference/finite-volume central scheme is used for the spatial discretization of the equations. {A discussion on the use of higher-order spatial discretization schemes is presented in Appendix C.}

The semi-discrete representation of the proposed ACDI PF model in Eq. \eqref{eq:ACDI} can be written as
\begin{equation}
\frac{\partial \phi}{\partial t} 
+ \frac{\hat{\Phi}_j\rvert_{(m+\frac{1}{2})} - \hat{\Phi}_j\rvert_{(m-\frac{1}{2})}}{\Delta x_j}
- \frac{\hat{A}_j\rvert_{(m+\frac{1}{2})} - \hat{A}_j\rvert_{(m-\frac{1}{2})}}{\Delta x_j}
= 0,
\label{eq:ACDI_discrete}
\end{equation}
where $\hat{\Phi}_j\rvert_{(m\pm{1}/{2})}$ and $\hat{A}_j\rvert_{(m\pm{1}/{2})}$ are the numerical fluxes for the convective and the interface-regularization (RHS) terms, respectively; the subscript $m$ denotes the grid-cell index; and $\Delta x_j$ represents the grid-cell size.
Now, adopting the flux-splitting procedure for the interface regularization terms from \citet{jain2022kinetic}, these numerical fluxes can be expressed as
\begin{equation}
    \begin{aligned}
    &\hat{\Phi}_j\rvert_{(m\pm\frac{1}{2})} = \xbar{\phi}^{(m\pm\frac{1}{2})} \xbar{u}_{j}^{(m\pm\frac{1}{2})},\\
    &\hat{A}_{j}\rvert_{(m\pm\frac{1}{2})} = \left\{\Gamma\left\{\frac{\epsilon}{\Delta x_j}\left(\Delta_j\phi\right) - \frac{1}{4} \left[1 - \tanh^2{\left(\frac{\xbar{\psi}^{(m\pm\frac{1}{2})}}{2\epsilon}\right)}\right]\xbar{\frac{\vec{\nabla} \psi}{|\vec{\nabla} \psi|}}^{(m\pm\frac{1}{2})}\right\}\right\},\\
    \end{aligned}
    \label{eq:discrete_flux}
\end{equation}
where the overbar denotes an arithmetic average of the quantity evaluated at $m$ and $m\pm1$; $\Delta_j \phi$ represents a difference along the $j$th coordinate as $\Delta_j \phi = \phi\rvert_{(m+1)}-\phi\rvert_m$. For the discrete representation of the consistent regularization terms in the mass, momentum, and energy equations, the same formulation in Eq. \eqref{eq:discrete_flux} should be used to maintain consistency. A skew-symmetric-like split scheme is used here, because these schemes result in reduced aliasing errors \citep{blaisdell1996effect,chow2003further,kennedy2008reduced} and are known to improve the conservation properties of the quadratic quantities \citep{kravchenko1997effect}. Moreover, the scheme in Eq. \eqref{eq:discrete_flux} is also responsible for the superior\textemdash a less restrictive\textemdash boundedness criterion in Section \ref{sec:bound} of the proposed PF model.

For the incompressible formulation presented in Section \ref{sec:incomp}, a staggered grid is used, where the velocities are stored at the staggered locations to avoid pressure checker boarding. 
The PF variable and the pressure are located at the cell center. {The staggered-grid formulation results in discrete conservation of total kinetic energy in the absence of dissipative mechanisms \citep{morinishi1998fully}.}

%
%
%

\section{Results}

In this section, the simulation results with the newly proposed ACDI method are presented. Two-dimensional interface advection test cases are presented in Sections \ref{sec:bubble_advection} and \ref{sec:drop_shear}, where the proposed ACDI method and the CDI methods are compared in terms of accuracy. More importantly, the effect of reducing $\epsilon$ on the convergence behavior and the accuracy of the methods are discussed in Section \ref{sec:effect_of_eps}, {the grid convergence study is presented in Section \ref{sec:grid_conv}, and the computational cost savings associated with using ACDI compared to CDI is discussed in Section \ref{sec:cost_save}}. This is followed by the incompressible hydrodynamics-coupled simulations in Section \ref{sec:HIT}, where the robustness of the proposed method is illustrated using the high-density-ratio droplet-laden isotropic turbulence simulation at infinite $Re$. 

{In all the simulations presented in this section, the mass of each phase is conserved, with the error on the order of machine precision; and $\phi$ is verified to be bounded between $0$ and $1$.}
To quantitatively compare the error of the proposed ACDI method with that of the CDI method, we use the error, $E$, defined as 
\begin{equation}
    E = \int_{\Omega}|\phi_f - \phi_o| dV
    = \sum_m \Big|\phi_{f}\rvert_m - \phi_{o}\rvert_m\Big| dV,
    \label{eq:error}
\end{equation}
where the subscripts $f$ and $o$ denote that the quantities are evaluated at the final and initial times, respectively.


\subsection{{Advection of a bubble/drop \label{sec:bubble_advection}}}

\begin{figure}
    \centering
    \includegraphics[width=\textwidth]{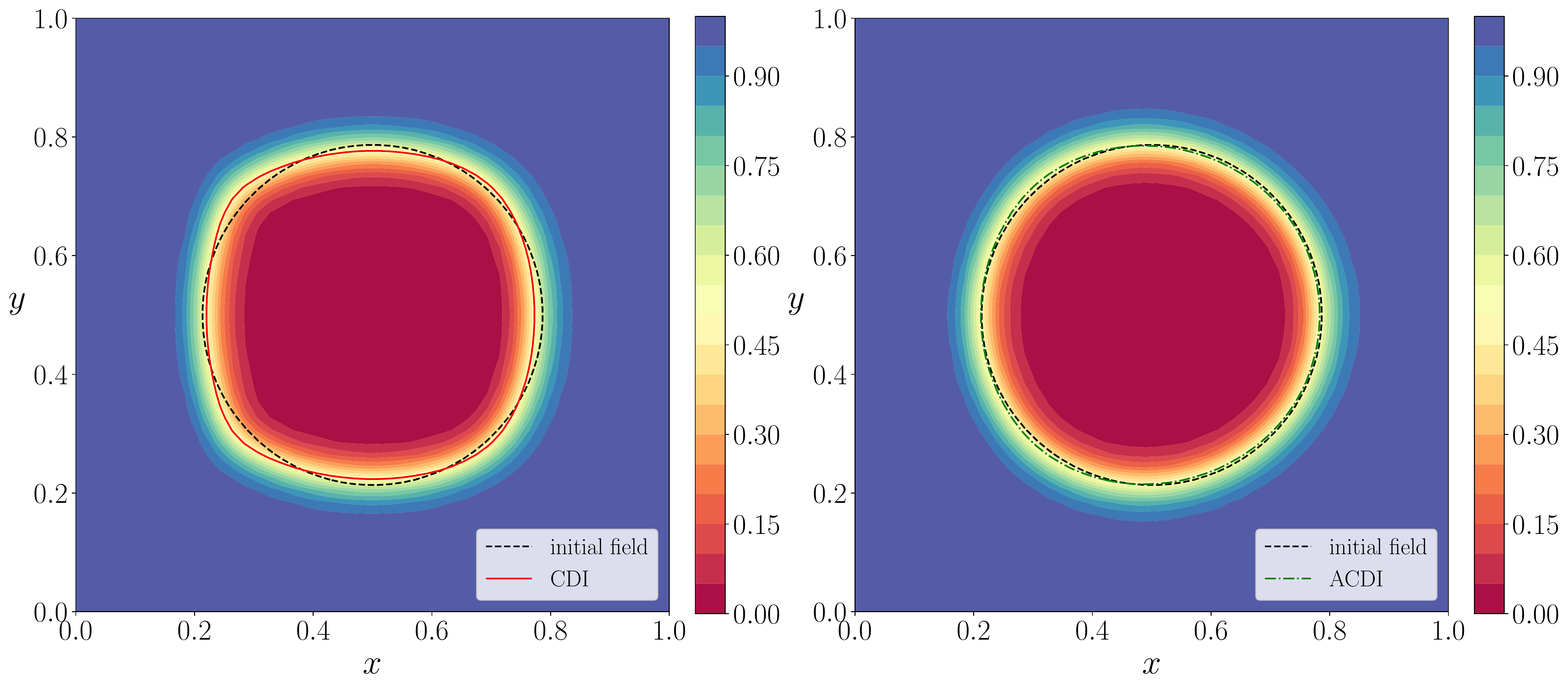}
    \caption{{Final shape of the translating drop/bubble computed using the proposed ACDI method in Eq. \eqref{eq:ACDI} and the CDI method in Eq. \eqref{eq:CDI}.}}
    \label{fig:bubble_advection}
\end{figure}

{In this test case, advection of a circular interface is considered to test the distortion of the interface that was seen with the CDI method in Figure \ref{fig:distort}. The setup is similar to the one presented in Section \ref{sec:intro}, where a two-dimensional circular bubble/drop is initialized at the center of the unit-sized periodic domain that is discretized into a $50\times50$ grid points. A uniform velocity field of $\vec{u}=5i$ is prescribed, $\Delta t=0.001$, and here the phase-field equation in Eq. \eqref{eq:ACDI} is solved (decoupled from hydrodynamics). The final and initial shape of the interface are presented in Figure \ref{fig:bubble_advection} along with the results from the CDI method [obtained by solving Eq. \eqref{eq:CDI}]. Unlike the CDI method, the proposed ACDI method maintains the interface shape for long-time integrations.}   
{To make a more quantitative comparison, the error $E$ computed using Eq. \eqref{eq:error}, for both the methods, is presented in Table \ref{tab:bubble_advection_error}. This shows that the proposed ACDI method is more accurate compared to the CDI method, although, error $E$ is not the best metric to highlight the significant improvement in shape that is seen in Figure \ref{fig:bubble_advection}.}
\begin{table}[H]
\centering
\begin{tabular}{@{}|l|l|@{}}
\toprule
CDI     & ACDI    \\ \midrule
0.01352 & 0.00576 \\ \bottomrule
\end{tabular}
\caption{{The error $E$ in the shape of the interface for the case of bubble/drop advection.}}
\label{tab:bubble_advection_error}
\end{table}


\subsection{Drop in a shear flow \label{sec:drop_shear}}
\begin{figure}
    \centering
    \includegraphics[width=0.75\textwidth]{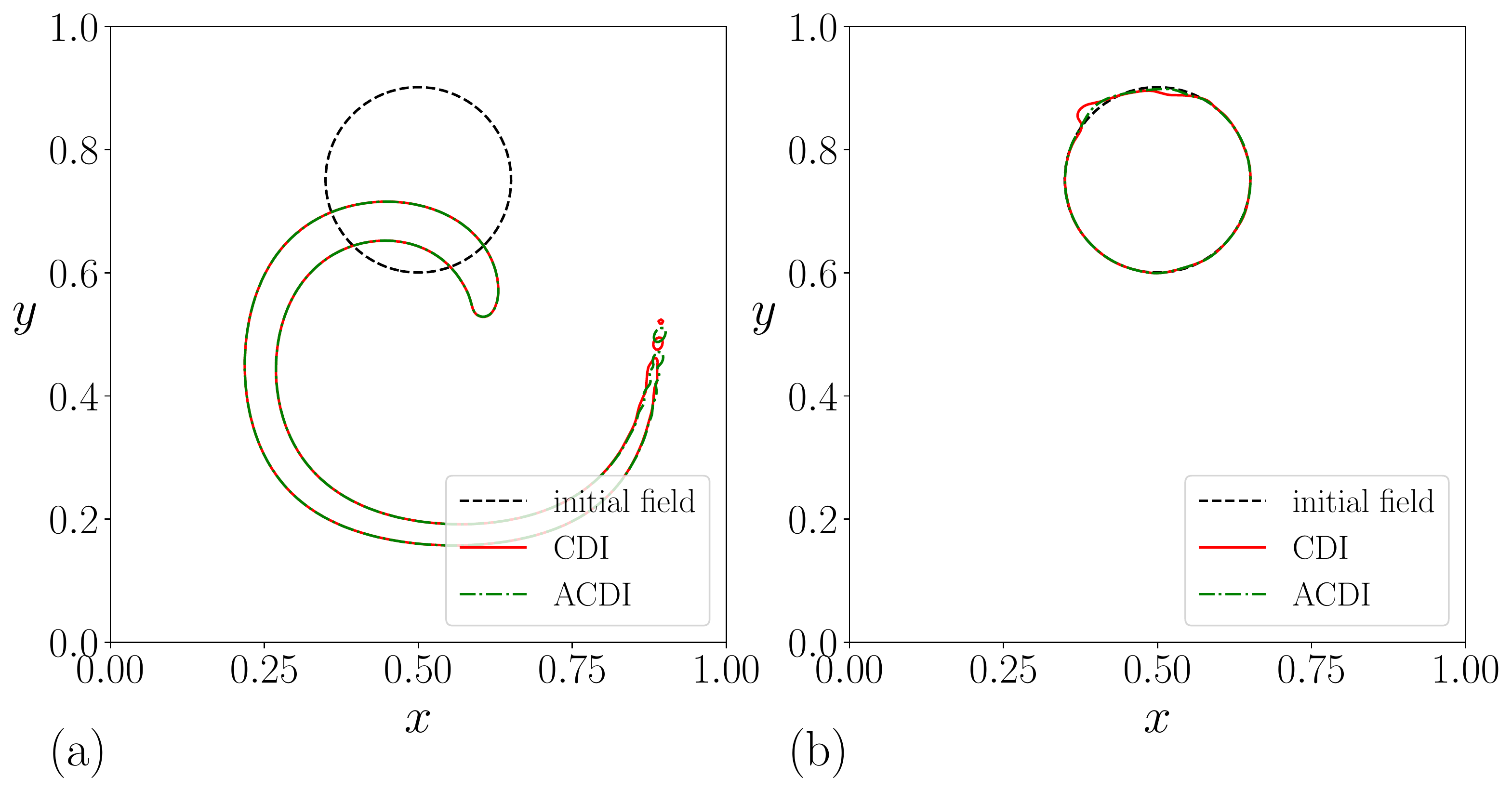}
    \caption{The shape of the drop at (a) half time of $t=2$ and (b) final time of $t=4$ in a shearing flow, computed in a domain discretized with a grid of size $256^2$.}
    \label{fig:drop_shear}
\end{figure}

This is a classical interface advection test case that was introduced by \citet{BELL1989257} and \citet{rider1998reconstructing} and has been extensively used in the literature \citep{Tryggvason2011} to assess the accuracy of the interface-capturing methods. In this test case, a circular drop/bubble of radius $R=0.15$ is initially placed in the unit-sized domain at $(0.5,0.75)$. A shearing velocity field given by
\begin{equation}
\begin{aligned}
    u = -\sin^2\left(\pi x\right)\sin(2\pi y)\cos\left(\frac{\pi t}{T}\right),\\ 
    v = \sin(2\pi x)\sin^2(\pi y)\cos\left(\frac{\pi t}{T}\right) 
\end{aligned}
\end{equation}
is prescribed, where $T=4$ is the time period of the flow; $t$ is the time coordinate; $x$ and $y$ are the spatial coordinates; and $u$ and $v$ are the velocity components along the $x$ and $y$ directions, respectively. {Here, $\Gamma=|\vec{u}|_{max}$ and $\epsilon=\Delta x$ are the interface parameters, and $\Delta t=2.5\times10^{-4}$ is the time-step size}. The drop undergoes a shearing deformation until $t=T/2=2$, and then the flow field is reversed with the hope that the initial drop shape is recovered at the final time of $t=T=4$. Figure \ref{fig:drop_shear} shows the shape of the drop at the half time of $t=2$ and the final time of $t=4$ for the ACDI and the CDI methods, computed on a grid of size $256^2$.  
The ACDI method is more accurate in recovering the circular shape of the drop compared with the CDI method. The error $E$ computed using Eq. \eqref{eq:error} for the ACDI method is 8.66$\times 10^{-4}$ and for the CDI method is 1.95$\times 10^{-3}$.

\subsubsection{Effect of choice of $\epsilon$ on the accuracy of the method \label{sec:effect_of_eps}}

As described in Sections \ref{sec:bound}-\ref{sec:cost}, it is important to use as small a value for $\epsilon$ as possible, while maintaining the boundedness of $\phi$, because this makes the resolved interface more sharp. Section \ref{sec:cost} showed that reducing the value of $\epsilon/\Delta x$ from $1$ to $0.5$ makes the simulation less expensive for the proposed ACDI method by increasing the maximum allowable time-step size, whereas reducing $\epsilon$ increases the cost for CDI method. In this section, the effect of varying the value of $\epsilon$ on the accuracy of the methods is evaluated.  

\begin{figure}
    \centering
    \includegraphics[width=0.75\textwidth]{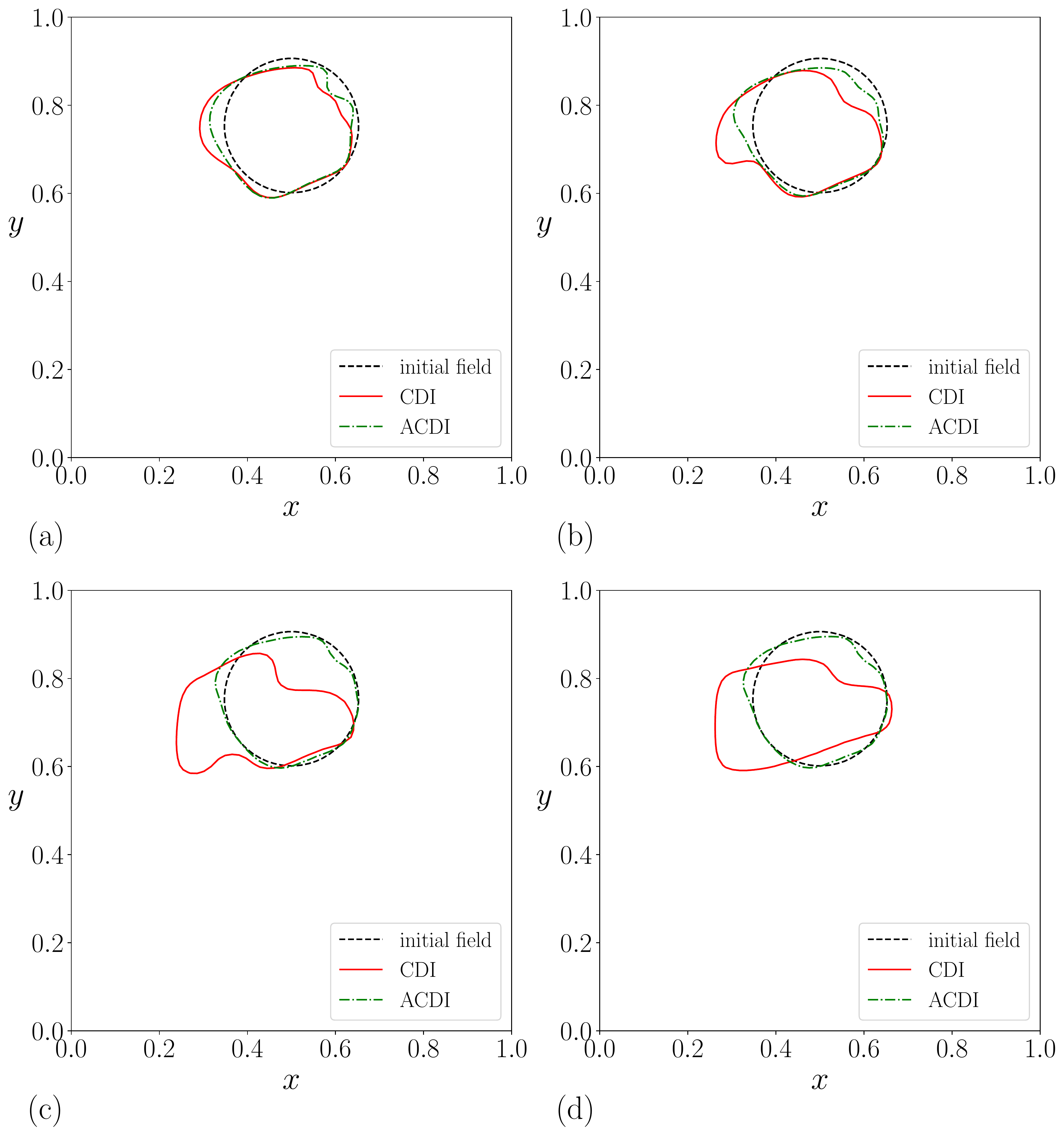}
    \caption{The final shape of the drop at $t=4$ for (a) $\epsilon/\Delta x=1$, (b) $\epsilon/\Delta x=0.75$, (c) $\epsilon/\Delta x=0.55$, and (d) $\epsilon/\Delta x=0.51$, in a shearing flow, computed in a domain discretized with a grid of size $64^2$.}
    \label{fig:eps_effect}
\end{figure}

\begin{figure}
    \centering
    \includegraphics[width=0.75\textwidth]{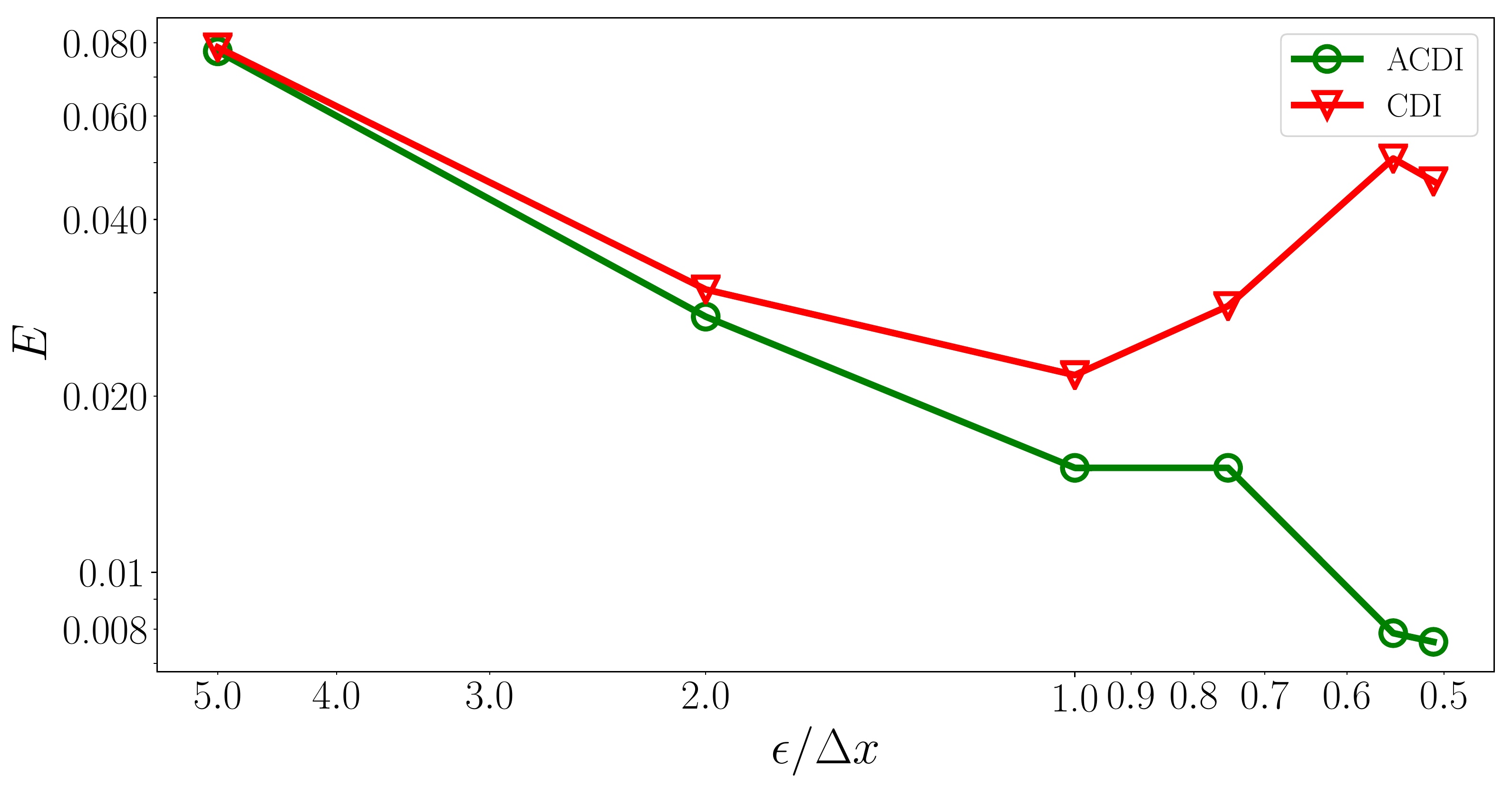}
    \caption{Comparison of the error, $E$, computed using Eq. \eqref{eq:error}, at the final time of $t=4$ for the CDI and ACDI methods for various values of $\epsilon$. The test case is the drop in a shear flow, computed in a domain discretized with a grid of size $64^2$.}
    \label{fig:eps_effect_error}
\end{figure}

Here, the test case of a drop in a shear flow from Section \ref{sec:drop_shear} is repeated, on a coarser grid of size $64^2$, {with $\Delta t=0.001$}, for various values of $\epsilon$, and the shape of the drop that is recovered at the final time of $t=4$ is shown in Figure \ref{fig:eps_effect}. For the proposed ACDI method, the final drop shape is more circular, and it approaches the ideal shape, as the value of $\epsilon$ is reduced. But for the CDI method, the final drop shape is more distorted as the value of $\epsilon$ is reduced.
Furthermore, the error $E$ is plotted against different values of $\epsilon$ in Figure \ref{fig:eps_effect_error} {and are also listed in Table \ref{tab:eps_effect}}. The error for the proposed ACDI method decreases as $\epsilon/\Delta x$ decreases, all the way up to $\epsilon/\Delta x = 0.5$. But the error for the CDI method increases as $\epsilon/\Delta x$ decreases for $\epsilon/\Delta x < 1$.  
Hence, the proposed ACDI method not only becomes less expensive as $\epsilon/\Delta x$ decreases, but also becomes more accurate; therefore, the method converges with decreasing $\epsilon/\Delta x$. The CDI method, in contrast, not only becomes more expensive as $\epsilon/\Delta x$ decreases but also becomes less accurate; therefore, the method diverges with decreasing $\epsilon/\Delta x$.  
%
\begin{table}[]
\centering
\begin{tabular}{@{}|c|ccc|ccl|@{}}
\toprule
     & \multicolumn{3}{c|}{CDI}                               & \multicolumn{3}{c|}{ACDI}                             \\ \midrule
$\epsilon/\Delta x$ &
  \multicolumn{1}{c|}{$\Gamma/|\vec{u}|_{max}$} &
  \multicolumn{1}{c|}{$E$} &
  \begin{tabular}[c]{@{}c@{}}relative\\ cost\end{tabular} &
  \multicolumn{1}{c|}{$\Gamma/|\vec{u}|_{max}$} &
  \multicolumn{1}{c|}{$E$} &
  \multicolumn{1}{c|}{\begin{tabular}[c]{@{}c@{}}relative\\ cost\end{tabular}} \\ \midrule
5    & \multicolumn{1}{c|}{1}  & \multicolumn{1}{c|}{0.078624} & 1    & \multicolumn{1}{c|}{1} & \multicolumn{1}{c|}{0.077414} & 1    \\ \midrule
2    & \multicolumn{1}{c|}{1}  & \multicolumn{1}{c|}{0.030380} & 1    & \multicolumn{1}{c|}{1} & \multicolumn{1}{c|}{0.027231} & 1    \\ \midrule
1    & \multicolumn{1}{c|}{1}  & \multicolumn{1}{c|}{0.021705} & 1    & \multicolumn{1}{c|}{1} & \multicolumn{1}{c|}{0.015077} & 1    \\ \midrule
0.75 & \multicolumn{1}{c|}{2}  & \multicolumn{1}{c|}{0.028467} & 1.5  & \multicolumn{1}{c|}{1} & \multicolumn{1}{c|}{0.015080} & 0.75 \\ \midrule
0.55 & \multicolumn{1}{c|}{10} & \multicolumn{1}{c|}{0.050817} & 5.5  & \multicolumn{1}{c|}{1} & \multicolumn{1}{c|}{0.007877} & 0.55 \\ \midrule
0.51 & \multicolumn{1}{c|}{50} & \multicolumn{1}{c|}{0.046484} & 25.5 & \multicolumn{1}{c|}{1} & \multicolumn{1}{c|}{0.007613} & 0.51 \\ \bottomrule
\end{tabular}
\caption{{Comparison of the error, $E$, computed using Eq. \eqref{eq:error}, at the final time of $t=4$ for the CDI and ACDI methods for various values of $\epsilon$. The test case is the drop in a shear flow, computed in a domain discretized with a grid of size $64^2$. Note that, for CDI, the value of $\Gamma/|\vec{u}|_{max}$ had to be increased for $\epsilon/\Delta x<1$ to maintain boundedness of $\phi$ according to Eq. \eqref{eq:CDI_bound}. Here, relative cost is calculated based on the relative number of time steps which is defined as $\Delta t_{reference}/\Delta t_{actual}=\Gamma/|\vec{u}|_{max} \times \epsilon/\Delta x$, and $\Gamma=|\vec{u}|_{max}$, $\epsilon/\Delta x=1$ are chosen as the reference.}}
\label{tab:eps_effect}
\end{table}


\subsubsection{{Effect of grid size on the accuracy of the method} \label{sec:grid_conv}}

\begin{figure}
    \centering
    \includegraphics[width=0.35\textwidth]{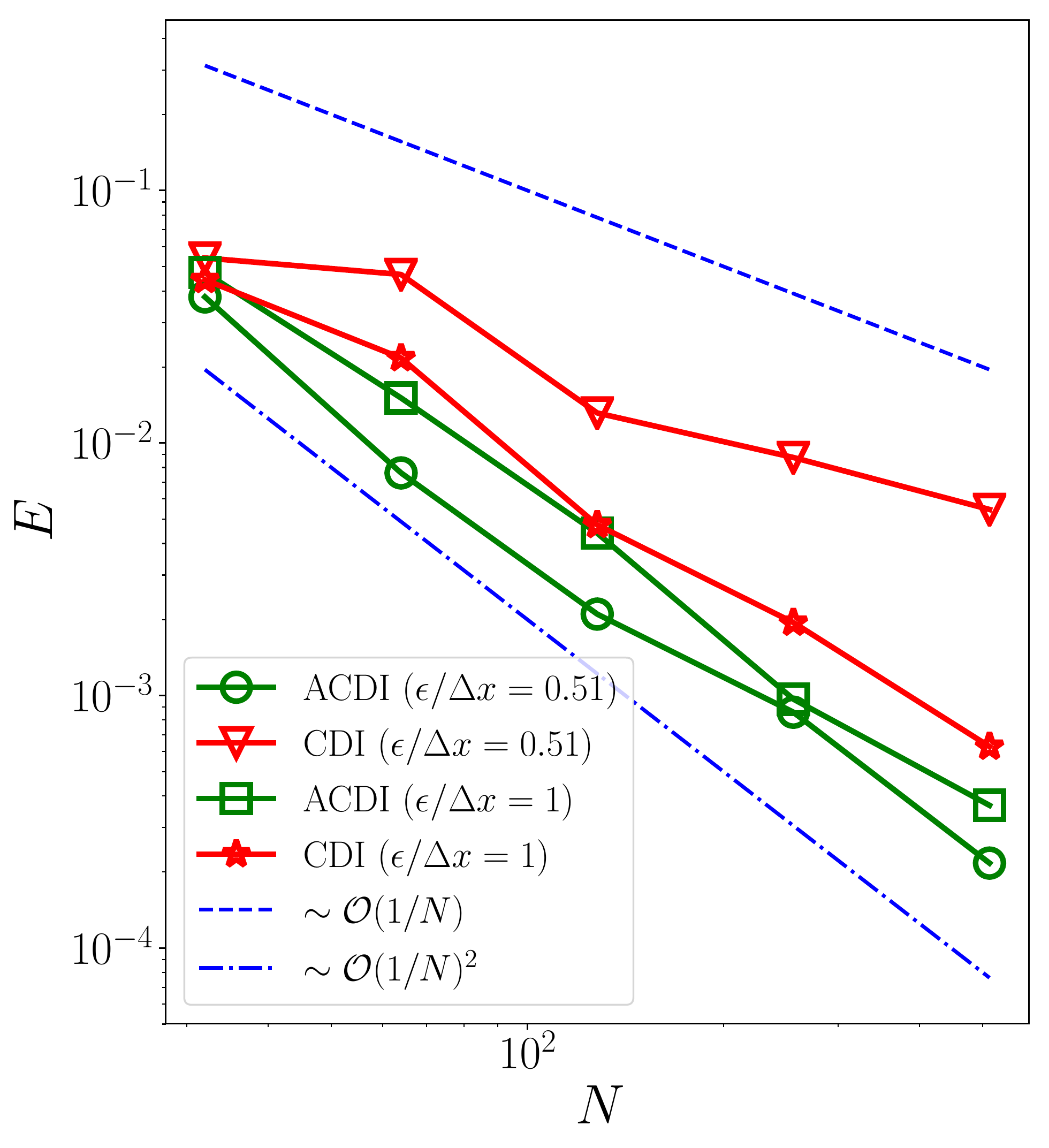}
    \caption{{Comparison of the error, $E$, computed using Eq. \eqref{eq:error}, at the final time of $t=4$ for the CDI and ACDI methods for various grid sizes $N$. The test case is the drop in a shear flow, computed in a domain discretized with a grid of size $N^2$.}}
    \label{fig:grid_conv}
\end{figure}

{In this section, the effect of varying the grid size on the accuracy of the ACDI and CDI methods is evaluated using the test case of a drop in a shear flow from Section \ref{sec:drop_shear}.
The domain is discretized using $N\times N$ grid points; and five different grids, $32^2$, $64^2$, $128^2$, $256^2$, and, $512^2$, were chosen to study the convergence of the error in the shape of the drop. Two different values of $\epsilon/\Delta x=1$ and $0.51$ were chosen. 
The computed error, $E$, on five different grids are listed in Table \ref{tab:grid_conv} along with the order of convergence. The error decreases with an increase in the number of grid points, with an order of convergence roughly between $1$ and $2$ for the ACDI method. The average order of convergence is $1.862$ for $\epsilon/\Delta x = 0.51$ and $1.752$ for {$\epsilon/\Delta x = 1$}. On the otherhand, for the CDI method, the average order of convergence is $0.8272$ for $\epsilon/\Delta x = 0.51$ and $1.539$ for $\epsilon/\Delta x = 1$. The error is also plotted against the grid size in Figure \ref{fig:grid_conv}; and the slopes of the individual line segments represent the local order of convergence that is listed in Table \ref{tab:grid_conv}. The ACDI method is also compared against other possible variations of the conservative second-order Allen-Cahn based phase-field models in Appendix D.} 

\begin{table}[]
\centering
\begin{tabular}{@{}|c|clcl|@{}}
\toprule
 &
  \multicolumn{4}{c|}{CDI} \\ \midrule
Grid &
  \multicolumn{1}{c|}{\begin{tabular}[c]{@{}c@{}}$E$ \\ ($\epsilon/\Delta x=0.51$)\end{tabular}} &
  \multicolumn{1}{l|}{order} &
  \multicolumn{1}{c|}{\begin{tabular}[c]{@{}c@{}}$E$ \\ ($\epsilon/\Delta x=1$)\end{tabular}} &
  order \\ \midrule
$32\times32$ &
  \multicolumn{1}{c|}{0.05403} &
  \multicolumn{1}{l|}{} &
  \multicolumn{1}{c|}{0.04428} &
   \\ \midrule
$64\times64$ &
  \multicolumn{1}{c|}{0.04648} &
  \multicolumn{1}{l|}{0.2172} &
  \multicolumn{1}{c|}{0.02171} &
  1.0458 \\ \midrule
$128\times128$ &
  \multicolumn{1}{c|}{0.01315} &
  \multicolumn{1}{l|}{1.822} &
  \multicolumn{1}{c|}{0.004741} &
  2.195 \\ \midrule
$256\times256$ &
  \multicolumn{1}{c|}{0.008764} &
  \multicolumn{1}{l|}{0.5854} &
  \multicolumn{1}{c|}{0.001946} &
  1.285 \\ \midrule
$512\times512$ &
  \multicolumn{1}{c|}{0.005455} &
  \multicolumn{1}{l|}{0.6840} &
  \multicolumn{1}{c|}{0.0006282} &
  1.631 \\ \midrule
\multicolumn{1}{|l|}{} &
  \multicolumn{4}{c|}{ACDI} \\ \midrule
Grid &
  \multicolumn{1}{c|}{\begin{tabular}[c]{@{}c@{}}$E$ \\ ($\epsilon/\Delta x=0.51$)\end{tabular}} &
  \multicolumn{1}{l|}{order} &
  \multicolumn{1}{c|}{\begin{tabular}[c]{@{}c@{}}$E$ \\ ($\epsilon/\Delta x=1$)\end{tabular}} &
  order \\ \midrule
$32\times32$ &
  \multicolumn{1}{c|}{0.03788} &
  \multicolumn{1}{l|}{} &
  \multicolumn{1}{c|}{0.04737} &
   \\ \midrule
$64\times64$ &
  \multicolumn{1}{c|}{0.007613} &
  \multicolumn{1}{l|}{2.315} &
  \multicolumn{1}{c|}{0.01508} &
  1.651 \\ \midrule
$128\times128$ &
  \multicolumn{1}{c|}{0.002102} &
  \multicolumn{1}{l|}{1.857} &
  \multicolumn{1}{c|}{0.004396} &
  1.778 \\ \midrule
$256\times256$ &
  \multicolumn{1}{c|}{0.0008591} &
  \multicolumn{1}{l|}{1.291} &
  \multicolumn{1}{c|}{0.0009681} &
  2.183 \\ \midrule
$512\times512$ &
  \multicolumn{1}{c|}{0.0002171} &
  \multicolumn{1}{l|}{1.984} &
  \multicolumn{1}{c|}{0.0003675} &
  1.397 \\ \bottomrule
\end{tabular}
\caption{{Comparison of the error, $E$, computed using Eq. \eqref{eq:error}, at the final time of $t=4$ for the CDI and ACDI methods. The test case is the drop in a shear flow. Note that, for CDI, the value of $\Gamma/|\vec{u}|_{max}=50$ is used for $\epsilon/\Delta x=0.51$ to maintain boundedness of $\phi$ according to Eq. \eqref{eq:CDI_bound}.}}
\label{tab:grid_conv}
\end{table}

\subsubsection{{Computational cost savings} \label{sec:cost_save}}

{In this section, the computational cost savings of the proposed ACDI method over the CDI method is calculated. This is done by estimating the increased spatial and temporal resolution required by the CDI method to yield similar accuracy as the ACDI method.} 

{From the study of the effect of $\epsilon$ on the accuracy of the methods in Section \ref{sec:effect_of_eps}, it was found that the CDI method is most accurate for $\epsilon/\Delta x\approx1$, and the ACDI method is most accurate for $\epsilon/\Delta x\approx 0.5$. 
Hence, now assuming the error of $0.0002171$ (Table \ref{tab:grid_conv}), obtained with the ACDI method on $512\times512$ grid for $\epsilon/\Delta x=0.51$, as the targeted error, the grid resolution required to achieve this error with the CDI method can be calculated. Assuming that the asymptotic convergence is already achieved at the resolution of $512\times512$, the error of $0.0006282$ obtained with the CDI method can be used to extrapolate the grid point requirement to achieve the targeted error of $0.0002171$. From Section \ref{fig:grid_conv}, The average order of convergence for the CDI method for $\epsilon/\Delta x=1$ is estimated to be $n=1.539$. Using the relation
\begin{equation}
    n=\frac{\log{(\frac{error_1}{error_2})}}{\log{(\frac{\Delta x_1}{\Delta x_2})}},
\end{equation}
the increased grid resolution required for the CDI method to yield similar accuracy as the ACDI method is calculated to be $N_{\Delta}=\Delta x_{ACDI}/\Delta x_{CDI}=1.99\approx2$. In three dimensions, the increased grid count is, therefore, $N_{\Delta}^3\approx2^3=8$.}

{In Section \ref{sec:cost}, the required time-step size as a function of $\epsilon$ for both CDI and ACDI are reported. Using $\epsilon/\Delta x=0.51$ for ACDI, the required time step size is $\Delta t_{ACDI}\sim \Delta x_{ACDI}/(|u|_{max}0.51)$, and using $\epsilon/\Delta x=1$ for CDI, the required time step size is $\Delta t_{ACDI}\sim \Delta x_{CDI}/|u|_{max}$. Therefore, the increased temporal resolution required for the CDI method to yield similar accuracy as the ACDI method is calculated to be $N_t=\Delta t_{ACDI}/\Delta t_{CDI}=1.99/0.51\approx4$.}

{Therefore, the total cost savings of using the ACDI method over the CDI method in three dimensions is about a factor of $N_{\Delta}^3 \times N_t=32$.}

\subsection{Droplet-laden isotropic turbulence \label{sec:HIT}}


In this section, the simulation of a droplet-laden isotropic turbulence is presented on coarse grids. It is well known that the coarse-grid simulation of a turbulent flow is a good test of the nonlinear stability and robustness of the method, which were previously studied, for single-phase flows, by \citet{honein2004higher}, \citet{mahesh2004numerical}, \citet{hou2005robust}, \citet{subbareddy2009fully}, and \citet{kuya2018kinetic}, and for two-phase flows by \citet{jain2022kinetic}. This is due to the lack of grid resolution to support dissipative mechanisms that stabilize the method. Here, high-density-ratio droplet-laden isotropic turbulence is simulated at finite and infinite $Re$. The infinite-$Re$ simulation corresponds to an inviscid case, where there are no dissipative mechanisms, and acts as a true test of robustness of the method. 


For the infinite-$Re$ simulation case, the proposed ACDI method is coupled with both the consistent momentum transport equation in Eq. \eqref{eq:momf}, which results in discrete kinetic energy conservation, and the standard momentum equation (inconsistent), which does not conserve kinetic energy. This is done to illustrate the importance of solving the consistent momentum transport equation.

The initial setup consists of a single spherical drop, of radius $R=1$, that is placed at the center of the triply periodic domain. 
The domain has dimensions of $2\pi\times2\pi \times 2\pi$, and is discretized into a grid of size $64\times64\times64$. {Here, $\Gamma=|\vec{u}|_{max}$ and $\epsilon=0.51\Delta x$ are the interface parameters, and $\Delta t=2.5\times10^{-3}$ is the time-step size.} The velocity field is initialized using the energy spectrum
\begin{equation}
    E(k)\propto k^4 \exp{\left[ -2 \left(\frac{k}{k_o}\right)^2 \right]},
\end{equation}
where $k$ is the wavenumber, and $k_o$ is the most energetic wavenumber. Here, $k_o$ is chosen to be equal to $4$, and hence, the initial Taylor microscale is set as $\lambda_o=2/k_o=0.5$.

The density of the droplet fluid is $\rho_1=1000$, and the density of the carrier fluid is $\rho_2=1$. The surface tension between the two fluids is set to zero; i.e., the Weber number is $\infty$. For the finite-$Re$ simulation, the dynamic viscosity of the two fluids is chosen to be $\mu_1=1.732$ and $\mu_2=1.732\times10^{-3}$, such that the initial Taylor-scale Reynolds number is $Re_{\lambda,o}=100$, where, the Taylor scale Reynolds number is defined as $Re_{\lambda}=u_{rms}\lambda/\nu$, where $u_{rms}=<u_i u_i>/3=(2/3)\int_{0}^{\infty} E(k) dk$ is the root-mean-square velocity fluctuation. For the infinite-$Re$ simulation, the dynamic viscosity of the two fluids is chosen to be $\mu_1=\mu_2=0$, such that $Re_{\lambda,o}=\infty$.

\begin{figure}
    \centering
    \includegraphics[width=0.65\textwidth]{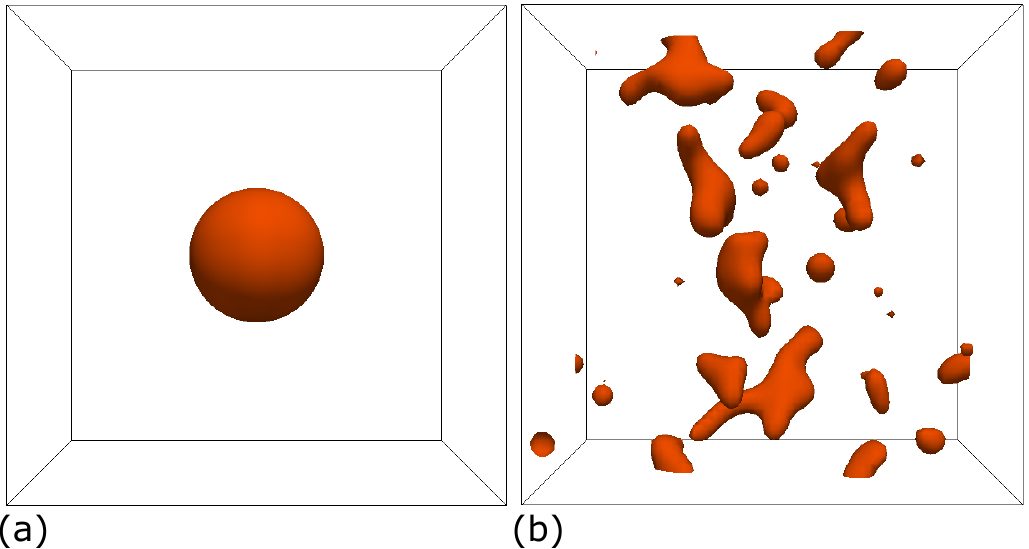}
    \caption{The snapshots from the droplet-laden isotropic turbulence simulation with the droplet-to-carrier fluid density ratio of $\rho_1/\rho_2=1000$, and the initial Taylor-scale Reynolds number of $Re_{\lambda,o}=100$, at (a) $t=0$ and (b) $t=5.7735$.}
    \label{fig:HIT_snap}
\end{figure}

The snapshots of the simulation results for the $Re_{\lambda,o}=100$ case at time $t=0$ and the final time of $t=5.7735$ are presented in Figure \ref{fig:HIT_snap}. Note that the breakup of the droplets and ligaments seen in this case is induced by the lack of grid resolution and, therefore, should not be considered physical. One needs to implement a subgrid-scale model to correctly predict the behavior of droplet breakup and the drop-size distribution. Here, we are instead interested in testing the robustness of the proposed ACDI method and the consistent momentum transport equation.  

\begin{figure}
    \centering
    \includegraphics[width=0.75\textwidth]{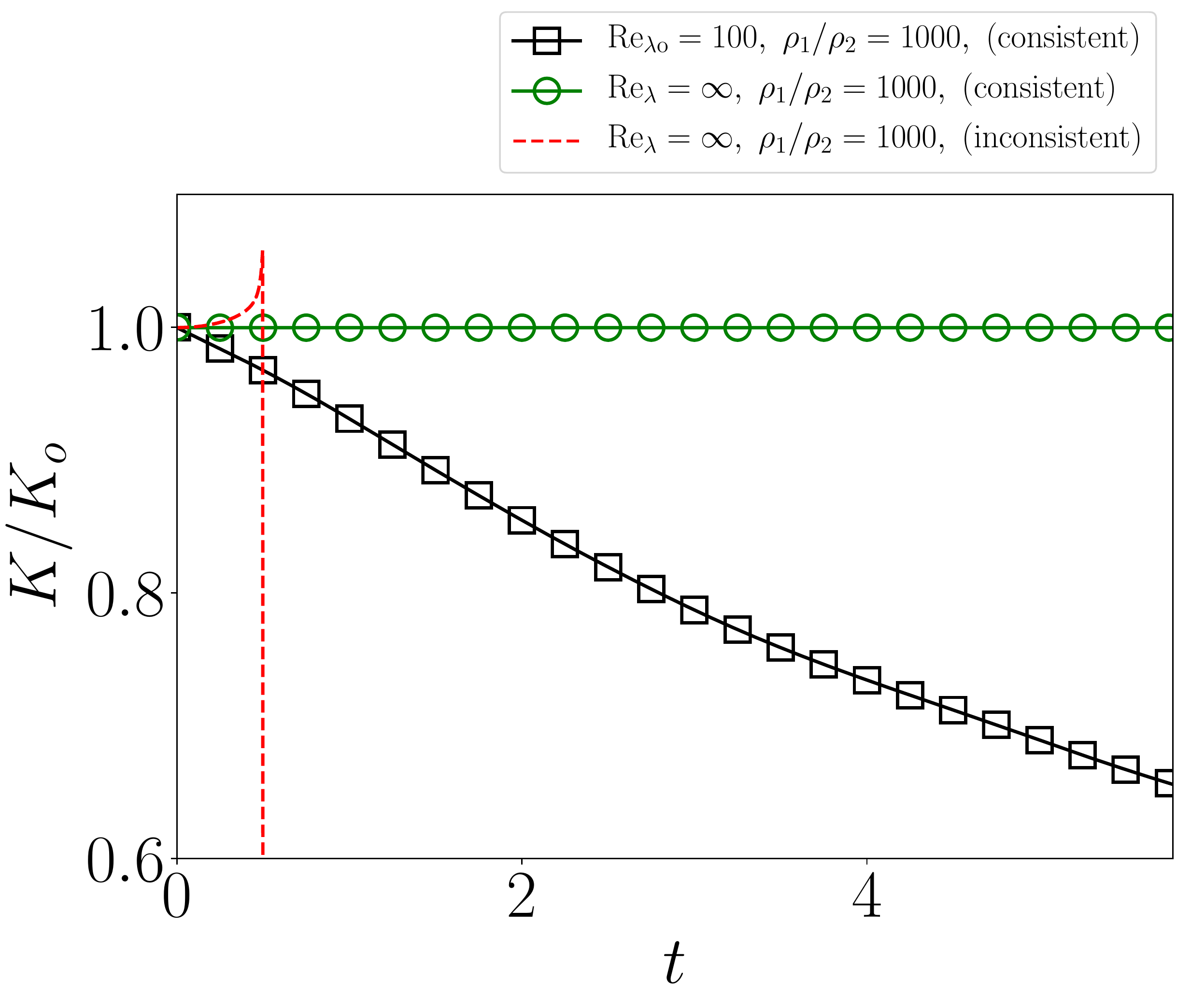}
    \caption{The evolution of total kinetic energy for the droplet-laden isotropic turbulence simulation for both finite and infinite $Re$. The consistent formulation is the PF model coupled with the proposed consistent momentum transport equation in Eq. \eqref{eq:momf}, and the inconsistent formulation is the PF model coupled with the standard momentum equation.}
    \label{fig:ke_evol}
\end{figure}

The time evolution of the total kinetic energy $K=\rho u_iu_i/2$, normalized by the initial value, for both finite- and infinite-$Re$ cases is shown in Figure \ref{fig:ke_evol}. For the finite-$Re$ case, the total kinetic energy of the system decays due to the viscous dissipation; and for the infinite-$Re$ case, the kinetic energy is conserved throughout the simulation (as expected), which results in stable numerical simulations. For the infinite-$Re$ case, results from the inconsistent momentum formulation are also shown in Figure \ref{fig:ke_evol} for comparison. Due to the inconsistent transport of momentum, the total kinetic energy is not conserved; therefore, the simulation diverges. 

\subsection{Spurious currents in a stationary drop/bubble}

In this section, a simple test case of a stationary drop/bubble is used to evaluate the accuracy in computing surface tension force with the proposed ACDI method. The CSF formulation involves computation of curvature, and the error in computing curvature results in spurious currents. Hence, the maximum magnitude of spurious currents can be used as a metric to estimate the accuracy in the implementation of surface tension force terms. 

\begin{figure}
    \centering
    \includegraphics[width=0.75\textwidth]{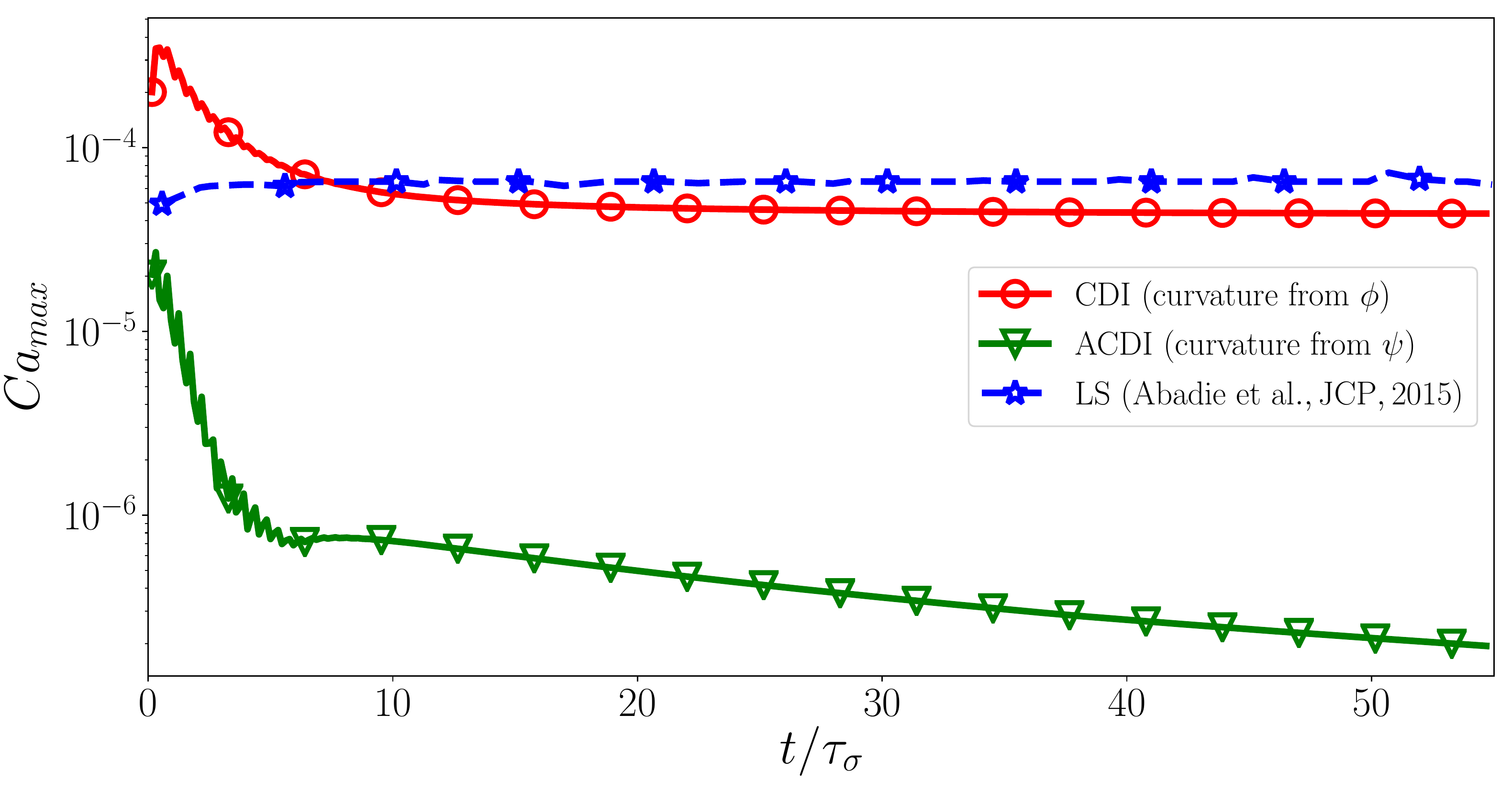}
    \caption{Evolution of the maximum non-dimensional spurious currents ($Ca_{max}$) in the domain with time. Time is normalized by the capillary time-scale $\tau_{\sigma}=\sqrt{\rho D^3/\sigma}$. LS denotes an LS method.}
    \label{fig:spurious}
\end{figure}

A two-dimensional stationary drop of radius $R=0.4$ is placed at the center of a square domain of unit size. {The domain is discretized into a grid of size $64^2$. Here, $\Gamma=|\vec{u}|_{max}$ and $\epsilon=0.75\Delta x$ are the interface parameters, and $\Delta t=0.001$ is the time-step size}. The density and viscosity of the droplet fluid and the surrounding fluid are set to be equal. The fluid properties are chosen such that the Laplace number is $La=\rho D \sigma/\mu^2=12000$. The maximum magnitude of the non-dimensional spurious currents ($Ca_{max}=\mu |\vec{u}|_{max}/\sigma$, where $Ca$ is the capillary number) in the domain is computed and plotted with time in Figure \ref{fig:spurious}. A significant reduction in the spurious currents, by over $2$ orders of magnitude, can be seen with the proposed ACDI method when the curvature is computed using Eq. \eqref{eq:new_kappa}, as opposed to Eq. \eqref{eq:old_kappa}. 


\begin{figure}
    \centering
    \includegraphics[width=\textwidth]{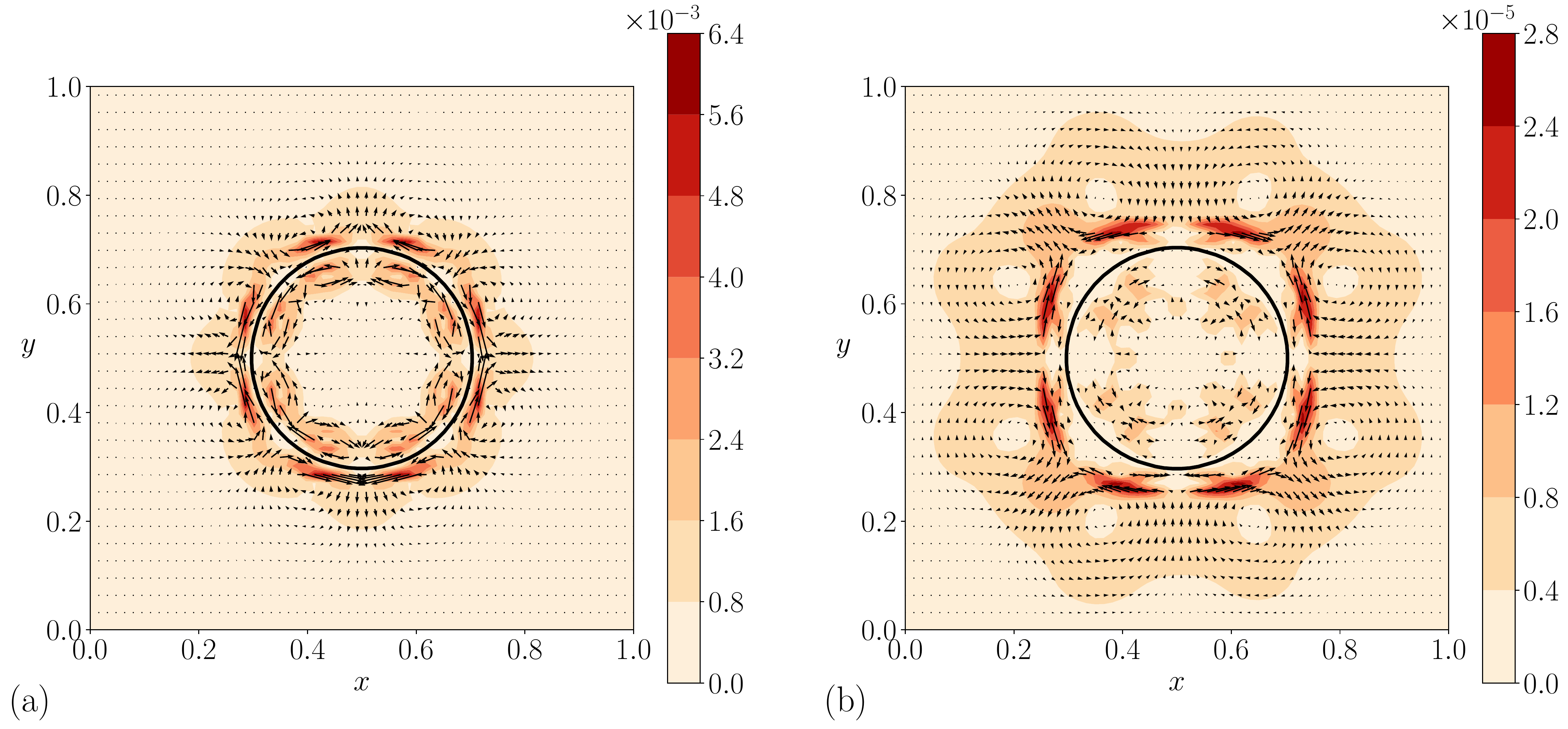}
    \caption{Form of the spurious currents around the interface at $t/\tau_{\sigma}\approx55$ with curvature computed using (a) $\phi$ [Eq. \eqref{eq:old_kappa}] and (b) $\psi$ [Eq. \eqref{eq:new_kappa}]. Here, the color represents the magnitude of the velocity in simulation units.}
    \label{fig:spurious_currents}
\end{figure}
    
Furthermore, to compare the present formulation with the LS formulation, the results from \citet{abadie2015combined} are also presented in Figure \ref{fig:spurious}. 
The curvature computation is very accurate in an LS method due to the use of a signed-distance function to compute interface normals. Although the improved curvature computation results in improving the accuracy of the surface-tension force, the well-balanced nature of the surface tension model gets destroyed due to the reinitialization step in the LS method \citep{abadie2015combined}. 
With the proposed ACDI method, the well-balanced nature of the surface tension model is retained, unlike the LS method, resulting in significantly lower magnitude of spurious currents compared with an LS method. 



\section{Conclusions} 

In this work, we proposed a novel PF model for the simulation of two-phase flows that is accurate, conservative, bounded, and robust. The proposed model conserves the mass of each of the phases and results in bounded transport of the volume fraction. We present results from the canonical test cases of a droplet advection and drop in a shear flow, showing an improvement in accuracy over the commonly used conservative PF method.

We also showed that the proposed method imposes lesser CFL restriction, and hence, is less expensive compared to previous methods. The proposed method also facilitates reformulation of the surface tension force calculation in terms of the new signed-distance variable. We showed that this improves the surface tension force calculation and significantly reduces the spurious currents at the interface.  
We further derived a consistent and conservative momentum transport equation for the proposed PF model 
that results in discrete conservation of kinetic energy, which is a sufficient condition for the numerical stability of incompressible flows in the absence of dissipative mechanisms. 

We discretize the proposed PF model using a flux-split conservative central scheme proposed by \citet{jain2022kinetic}, which reduces the aliasing error and is also responsible for the superior boundedness property of the model. 
To illustrate the robustness of the method in simulating high-density-ratio turbulent two-phase flows, we present the numerical simulations of droplet-laden isotropic turbulence at infinite Reynolds numbers for a density ratio of $1000$.

\section*{Acknowledgments} 

S. S. J. acknowledges financial support from the Franklin P. and Caroline M. Johnson Graduate Fellowship. {S. S. J. also acknowledges fruitful discussions with Parviz Moin throughout this study.} A preliminary version of this work has been published as a technical report in the annual publication of the Center for Turbulence Research \citep{jain2021acdi} and is available online\footnote{http://web.stanford.edu/group/ctr/ResBriefs/2021/15\_Jain.pdf}. {S. S. J. also acknowledges the comments by Shahab Mirjalili on this technical report. S. S. J. is thankful for the comments by the reviewers that helped improve this work.}



\section*{Appendix A: Reduced error in the computation of curvature}

The improvement in the computation of surface tension forces proposed in Section \ref{sec:surface} is due to the reduced truncation errors in the computation of $\kappa$ using Eq. \eqref{eq:new_kappa} as opposed to Eq. \eqref{eq:old_kappa}. This reduction in truncation errors can be easily seen as follows. 

Assuming 1D for simplicity, and discretizing $\kappa$ as
\begin{equation}
    k_m = \frac{n\rvert_{(m+\frac{1}{2})} - n\rvert_{(m-\frac{1}{2})}}{\Delta x}
\end{equation}
where $n$ is the 1D normal vector along $x$ direction. For the formulation in Eq. \eqref{eq:old_kappa} that uses $\phi$ to compute $\kappa$, the error in the discrete computation of $n$ is proportional to the truncation error in the computation of $\phi' \rvert_{(m+\frac{1}{2})}$, which can be expressed as
\begin{equation}
    \varepsilon_{\phi} = \phi' \rvert_{(m+\frac{1}{2})} - \frac{\phi\rvert_{(m+1)} - \phi_m}{\Delta x} = - \frac{(\Delta x)^2}{24} \phi '''(x\rvert_{(m+\frac{1}{2})}),
\end{equation}
where $'$ denotes a derivative here. If $\phi \sim \tanh{x}$, then $\phi''' \sim (4 \tanh^2{x}\ \text{sech}^2\ {x} - 2 \text{sech}^4\ {x})$, which is nonzero at $x=0$ (interface location). Note that the odd derivatives of $\phi$ are non-zero at $x=0$ and the value of these derivatives at $x=0$ increases with the increase in the order of the derivative. On the other hand, for the formulation in Eq. \eqref{eq:new_kappa} that uses $\psi$ to compute $\kappa$, the error in the discrete computation of $n$ is proportional to the truncation error in the computation of $\psi' \rvert_{(m+\frac{1}{2})}$, which can be expressed as
\begin{equation}
    \varepsilon_{\psi} = - \frac{(\Delta x)^2}{24} \psi '''(x\rvert_{(m+\frac{1}{2})}).
\end{equation}
If $\psi \sim x$, then $\psi''' \sim 0$. Therefore, the computation of $\kappa$ using the formulation in Eq. \eqref{eq:new_kappa} is much more accurate than the computation of $\kappa$ using the formulation in Eq. \eqref{eq:old_kappa}.
Note, however, $\phi$ is not exactly equal to $\tanh{x}$ and $\psi$ is not exactly equal to $x$, unless the interface is in perfect equilibrium. Hence, this is an approximate analysis and the computation of $\kappa$ is, therefore, not exact using the relation in Eq. \eqref{eq:new_kappa}, albeit being more accurate.

\section*{{Appendix B: Compressible flows}}

{The extension of the proposed phase-field model for compressible flows follows naturally from the recent works of \citet{jain2020conservative,jain2021assessment}. The interface regularization in the five-equation model \citep{jain2020conservative} or the four-equation model \citep{jain2021assessment} can be replaced with the regularization term proposed in this work in Eq. \eqref{eq:regularization}.
With this, the consistent transport equations for the mass, momentum, and energy can be derived for the proposed ACDI method, and the coupled system of equations can be written as
\begin{equation}
\frac{\partial \phi_1}{\partial t} + \vec{\nabla}\cdot(\vec{u}\phi_1) = \phi_1(\vec{\nabla}\cdot\vec{u})+\zeta_1(\vec{\nabla}\cdot\vec{u})+\vec{\nabla}\cdot\vec{a}_1,
\label{eq:volumefcomp}    
\end{equation}
\begin{equation}
\frac{\partial m_l}{\partial t} + \vec{\nabla}\cdot(\vec{u} m_l) = \vec{\nabla}\cdot \vec{R}_l, \hspace{0.5cm} l=1,2,
\label{eq:massfcomp}
\end{equation}
\begin{equation}
\frac{\partial \rho\vec{u}}{\partial t} + \vec{\nabla}\cdot(\rho \vec{u} \otimes \vec{u} + p \mathds{1}) = \vec{\nabla}\cdot\doubleunderline\tau + \vec{\nabla}\cdot(\vec{f}\otimes\vec{u}) + \sigma \kappa \vec{\nabla} \phi_1 + \rho \vec{g},
\label{eq:momfcomp}
\end{equation}
\begin{equation}
\frac{\partial E}{\partial t} + \vec{\nabla}\cdot\{(E + p)\vec{u}\} = \vec{\nabla}\cdot(\doubleunderline\tau\cdot\vec{u}) + \vec{\nabla}\cdot(\vec{f}k) + \sum_{l=1}^2 \vec{\nabla}\cdot{(\rho_l h_l \vec{a}_l )} + \sigma \kappa \vec{u}\cdot\vec{\nabla} \phi_1 + \rho \vec{g}\cdot \vec{u},
\label{eq:energyfcomp}
\end{equation}
\begin{equation}
p = f(\rho e, \phi_l, \alpha_l, \beta_l, \gamma_l,...),
\label{eq:pressurefcomp}
\end{equation}
where $m_l=\rho_l \phi_l$ is the partial density of phase $l$; $\vec{R}_l=\rho_l \vec{a}_l$ is the consistent mass flux; $h_l=e_l+p_l/\rho_l$ is the specific enthalpy of phase $l$; $E=\rho(e+k)$ is the total energy of the mixture per unit volume; $e$ is the specific mixture internal energy; and the function $\zeta_1$ is given by
\begin{equation}
    \zeta_1 = \frac{\rho_2 c_2^2 - \rho_1 c_1^2}{\frac{\rho_1 c_1^2}{\phi_1} + \frac{\rho_2 c_2^2}{\phi_2}},
\end{equation}
where $c_l$ is the speed of sound for phase $l$. The system of equations [Eqs. \eqref{eq:volumefcomp}-\eqref{eq:energyfcomp}] is closed using a generalized mixture equation of state given in Eq. \eqref{eq:pressuref}, where $\alpha_l$, $\beta_l$, and $\gamma_l$ are the material parameters for phase $l$.}

\section*{{Appendix C: Higher-order spatial schemes}}
{Throughout this work, a flux-split second-order central scheme from \citet{jain2022kinetic} has been used. However, an appropriate higher-order scheme can also be used with the proposed ACDI model. For example, following the generalization of flux-splitting procedure for higher-order schemes in \citet{ducros2000high} and \citet{pirozzoli2010generalized}, a fourth-order flux-split central scheme for the sharpening and convective terms that will still result in maintaining the boundedness of $\phi$ can be written as}
\begin{equation}
\footnotesize
    \begin{aligned}
    &{\hat{\Phi}_j\rvert_{(m\pm\frac{1}{2})} = \sum_{i=1}^2 \left[ a_i \left(\frac{\phi\rvert_m + \phi\rvert_{(m\pm i)}}{2}\right) \left(\frac{u_j\rvert_m + u_j\rvert_{(m\pm i)}}{2}\right)\right],}\\
    &{\hat{A}_{j}\rvert_{(m\pm\frac{1}{2})} = \left\{\Gamma\left\{\frac{\epsilon}{\Delta x_j}\left(\Delta_j\phi\right) -
    \sum_{i=1}^2 \left\{ a_i \frac{1}{4} \left[1 - \tanh^2{\left(\frac{\psi\rvert_m + \psi\rvert_{(m\pm i)}}{4\epsilon}\right)}\right]\left(\frac{\left.\frac{\vec{\nabla} \psi}{|\vec{\nabla} \psi|}\right|_m + \left.\frac{\vec{\nabla} \psi}{|\vec{\nabla} \psi|}\right|_{(m\pm i)}}{2} \right)\right\}\right\}\right\},}\\
    \end{aligned}
    \label{eq:discrete_flux_high_order}
\end{equation}
{where $a_1=4/3$ and $a_2=-1/6$ are the coefficients.
Similarly, any arbitrarily high-order flux-split central scheme can be constructed. 
Note that, unlike the proposed ACDI model, higher-order schemes for the CDI model will not result in maintaining the boundedness of $\phi$.} 

To evaluate the accuracy of the ACDI model with a higher-order scheme, the test case of a drop in a shear flow from Section \ref{sec:drop_shear} is repeated, on grids of sizes $64^2$ and $256^2$, using the second-order scheme in Eq. \eqref{eq:discrete_flux} and the fourth-order scheme in Eq. \eqref{eq:discrete_flux_high_order}, for various values of $\epsilon$. 
Here, a value of $\Gamma/|u|_{max}=1.5, 2$ were used for $\epsilon/\Delta x=0.55,0.51$, respectively, for the fourth-order scheme. 
For a higher-order scheme, 
the boundedness criterion is $\Gamma\ge C|\vec{u}|_{max}\ \mathrm{and}\ \epsilon > 0.5 \Delta x$, where $C$ is a value greater than 1. For a fourth-order scheme, $C$ was found, from inspection, to be approximately $2$.

\begin{figure}
    \centering
    \includegraphics[width=\textwidth]{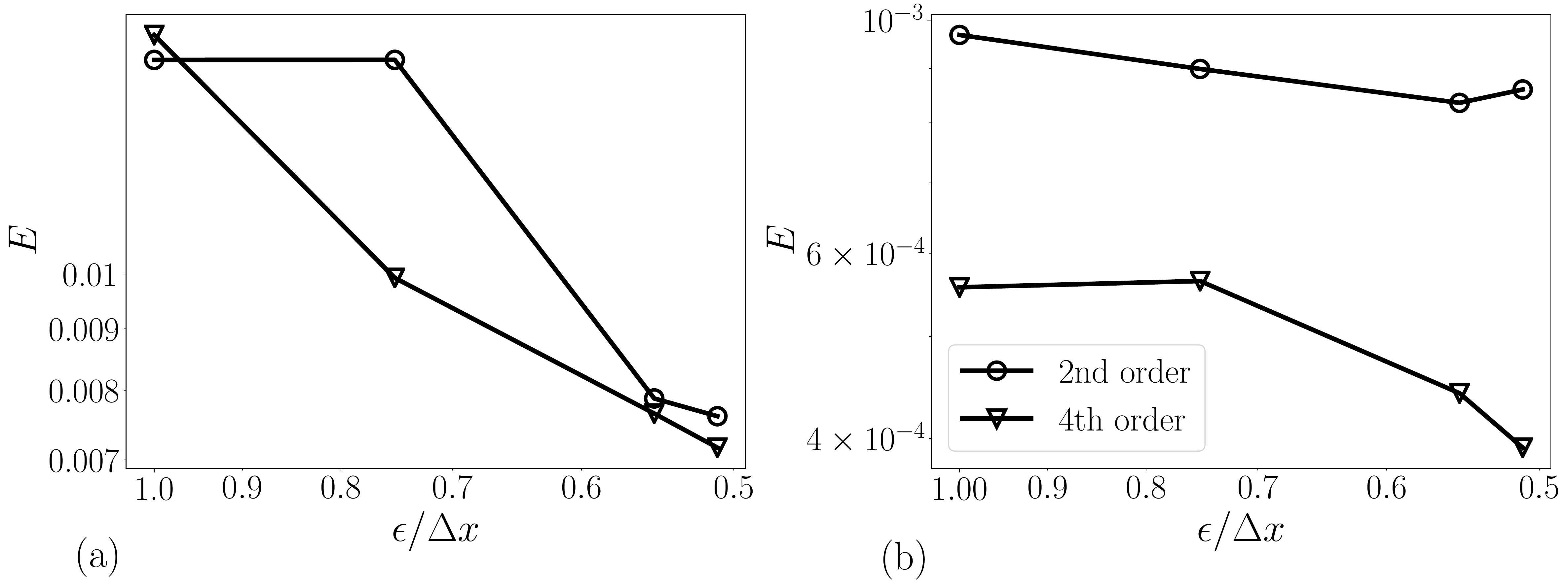}
    \caption{The error, $E$, computed using Eq. \eqref{eq:error}, at the final time of $t=4$ for the drop in a shear flow case. Here, the figures represent the error calculated in domain that is discretized with the grids of size (a) $64^2$ (coarse simulation), and (b) $256^2$ (refined simulation).}
    \label{fig:high_order}
\end{figure}

The final drop shape at $t=4$ is not shown here, but the error $E$ is plotted against different values of $\epsilon$ in Figure \ref{fig:high_order}. On the coarser grid of size $64^2$, both the second-order and fourth-order schemes have relatively similar accuracy for all values of $\epsilon$ considered. But on the refined grid of size $256^2$, the fourth-order scheme yields more accurate results compared to the second-order scheme. This is because the asymptotic convergence of the fourth-order scheme is only achieved in the refined-grid limit. 
Furthermore, to make a true comparison of the results with the second-order scheme and the fourth-order scheme, the average computational cost of the simulations, for the four simulations in Figure \ref{fig:high_order} (b) are computed and listed in Table \ref{tab:high_order}. The computational cost of the simulations with the fourth-order scheme is also higher compared to the second-order scheme. Hence, it is a trade-off between cost and accuracy.

\begin{table}[]
\centering
\begin{tabular}{@{}|c|c|c|@{}}
\toprule
                                                             & 2nd order & 4th order \\ \midrule
\begin{tabular}[c]{@{}c@{}}Average\\ time taken\end{tabular} & 129.1 s   & 222 s     \\ \bottomrule
\end{tabular}
\caption{Average time taken, in seconds, for the drop in a shear flow case in Figure \ref{fig:high_order} (b). The simulation was run on 256 Intel Xeon CPU cores.}
\label{tab:high_order}
\end{table}

With these observations, the best-practice recommendation for simulation of two-phase flows can be stated as follows: for a coarse-grid large-eddy simulation (LES), a second-order scheme is sufficient; but for a refined-grid direct numerical simulation (DNS), a higher-order scheme is viable, which results in more accurate simulations albeit at higher costs.



\section*{{Appendix D: Variations of the phase-field model}}


{In this section, the proposed ACDI model is compared against the other possible variations of the conservative second-order Allen-Cahn-based phase-field model that one could construct. These models can be written as
\begin{equation}
(model\ A) \hspace{2em} \frac{\partial \phi}{\partial t} + \vec{\nabla}\cdot(\vec{u}\phi) = \vec{\nabla}\cdot\left\{\Gamma\left\{\epsilon\vec{\nabla}\phi - \frac{1}{4} \left[1 - \tanh^2{\left(\frac{\psi}{2\epsilon}\right)}\right]\frac{\vec{\nabla} \phi}{|\vec{\nabla} \phi|}\right\}\right\},
\label{eq:modelA}
\end{equation}
\begin{equation}
(model\ B) \hspace{2em} \frac{\partial \phi}{\partial t} + \vec{\nabla}\cdot(\vec{u}\phi) = \vec{\nabla}\cdot\left\{\Gamma\left\{\frac{1}{4} \left[1 - \tanh^2{\left(\frac{\psi}{2\epsilon}\right)}\right]\left(\vec{\nabla}\psi - \frac{\vec{\nabla} \psi}{|\vec{\nabla} \psi|}\right)\right\}\right\}.
\label{eq:modelB}
\end{equation}
The difference between \textit{model A} in Eq. \eqref{eq:modelA} and the ACDI model is that the normal vector, in the sharpening term, is computed in terms of $\phi$. For \textit{model B} in Eq. \eqref{eq:modelB}, the diffusion term is also expressed in terms of $\psi$. 
The comparison of error $E$, computed using Eq. \eqref{eq:error}, is presented in Table \ref{tab:compare_PFs} for the drop in a shear flow case from Section \ref{sec:drop_shear}. Here, $\Gamma=|\vec{u}|_{max}$, and $\epsilon/\Delta x=1$ are chosen as the parameters. 
}
\begin{table}[]
\centering
\begin{tabular}{@{}|c|c|c|c|c|c|@{}}
\toprule
$E$     & $32\times32$ & $64\times64$ & $128\times128$ & $256\times256$ & $512\times512$ \\ \midrule
ACDI    & 0.04737      & 0.01508      & 0.004396       & 0.0009681      & 0.0003675      \\ \midrule
CDI     & 0.04428      & 0.02171      & 0.004741       & 0.001946       & 0.0006282      \\ \midrule
model A & 0.04678      & 0.02152      & 0.005297       & 0.001844       & 0.0005914      \\ \midrule
model B & 0.04637      & 0.01530      & 0.003821       & unstable       & unstable       \\ \bottomrule
\end{tabular}
\caption{{Comparison of the error, $E$, computed using Eq. \eqref{eq:error}, at the final time of $t=4$ for the CDI and ACDI methods along with the other possible variations of the conservative second-order Allen-Cahn-based phase-field model. The test case is the drop in a shear flow.}}
\label{tab:compare_PFs}
\end{table}

\bibliographystyle{model1-num-names}
\bibliography{two_phase}

\end{document}